\documentclass[a4paper,11pt]{article}
\usepackage{amsmath}
\pdfoutput=1
\usepackage{jheppub}
\usepackage{slashed}

\def\ie{{\it i.e.\;}}
\def\eg{{\it e.g.\;}}
\def\ope{\mathcal{O}}

\def\anj{\langle n_\text{jets} \rangle}
\def\eq#1{Eq.~(\ref{#1})}
\def\fig#1{Fig.~\ref{#1}}
\def\qqquad{\qquad \qquad}
\def\nj{n_\text{jets}}

\title{Scaling Patterns for QCD Jets}

\abstract{Jet emission at hadron colliders follows simple scaling
  patterns. Based on perturbative QCD we derive Poisson and staircase
  scaling for final state as well as initial state radiation. Parton
  density effects enhance staircase scaling at low multiplicities. We
  propose experimental tests of our theoretical findings in Z+jets 
  and QCD gap jets production, based on minor additions to current 
  LHC analyses.}

\author[a]{Erik Gerwick,}
\author[b]{Tilman Plehn,}
\author[a]{Steffen Schumann,}
\author[b]{and Peter Schichtel}

\affiliation[a]{II. Physikalisches Institut, Universit\"at G\"ottingen, 
                Germany}
\affiliation[b]{Institut f\"ur Theoretische Physik, Universit\"at Heidelberg,
                Germany}

\emailAdd{erik.gerwick@phys.uni-goettingen.de}
\emailAdd{plehn@uni-heidelberg.de}
\emailAdd{steffen.schumann@phys.uni-goettingen.de}
\emailAdd{p.schichtel@thphys.uni-heidelberg.de}
\begin{document}
\maketitle
\newpage

\section{Multi-jet rates}
\label{sec:intro}

Multi-jets final states are ubiquitous at hadron colliders. QCD
jet radiation mostly off the initial state partons has huge impact on
almost every LHC analysis. For example in Higgs searches an accurate
description of the jet recoil against the Higgs boson allows for an
efficient rejection of many backgrounds~\cite{us_prl}. In top pair
production and single top production the identification of QCD recoil
jets on the one hand and decay jets on the other hand is one of the
limiting factors in precision analyses~\cite{top_pairs,top_single}. New physics
searches largely rely on hard decay jets of new strongly interacting
particles, which makes them vulnerable in the case of soft decay jets
hidden in the QCD jet
activity~\cite{us_auto,mkraemer,review}. Understanding the jet
multiplicity and the jet spectra from QCD radiation is a core
ingredient to improving any of these analyses.\medskip

Even though jet radiation seems to follow simple
patterns~\cite{scaling_orig,us_photon}, theoretical predictions for
multi-jet observables in perturbative QCD are challenging.  It is
clear that the radiation of relatively hard quarks and gluons is a
direct consequence of the hierarchy between the large proton-proton
collider energy and the typical electroweak scale of the partonic
interaction~\cite{basics_pqcd,esw}. Numerically, we can combine the
QCD parton shower with hard matrix element calculations, to predict jet
radiation patterns over wide phase space
regions~\cite{ckkw,mlm,lecture}. Analytically, Sudakov factors and
generating functionals can be used to describe QCD jet
radiation~\cite{had_gf}.  A careful comparison of analytical and
numerical approaches to LHC data would allow us to determine the
strengths and limitations of the underlying theoretical concepts:
fixed-order perturbation theory, parton showers, and resummation based
on generating functionals.\bigskip

In fixed order perturbation theory leading-order jet rates are
available for effectively arbitrarily high 
multiplicities~\cite{mg5,comix}.  The number of jets we can consider 
is limited only by computing power. However, leading order predictions 
suffer significant shortcomings when it comes to precision. The 
renormalization scale dependence as a measure for the theoretical 
uncertainty grows with each power of $\alpha_s$ when we
add a final state jet. High powers of the logarithmic scale dependence
$(\log \mu_R)^n$ mean that such cross section predictions can only be
considered an order-of-magnitude estimate.
In addition, in the presence of phase space constraints large
logarithms spoil the convergence of fixed order perturbation theory.
For example jet vetoes will induce sizeable logarithms of ratios of
leading jet $p_T$ or masses ($m_H$ or $m_Z$) to additional
resolved jets (experimentally relevant down to 
$20-30$~GeV)~\cite{neubert,banfi}.

Next-to-leading order computations ameliorate the scale dependence and
capture an additional logarithm.  However, we are still limited in 
the available final-state multiplicity, \eg pure jets are 
available for $\nj \le 4$~\cite{pure_jets}, $\nj \le 4$ in association 
with $W/Z$ bosons~\cite{bh_wjets,bh_zjets}, for $t\bar{t}$ production and
Higgs production in gluon fusion NLO corrections are known for 
$\nj\le 2$ \cite{ttbar,gg_higgs}.  However, over the past few years this 
field has progressed enormously. As a consequence for Standard Model 
processes a similar level of automation as for leading-order 
calculations is within reach \cite{nlo_auto}. An approach applicable 
for general New Physics extensions, though limited to $2\to2$ processes, 
has been presented in~\cite{mad_golem}. 

Although NLO calculations contain one additional power of enhanced
logarithms, this might not be sufficient for high jet
multiplicities. At NNLO, although there has been an enormous amount
of recent development~\cite{NNLO}, the number of fully differential
calculations is limited, and an automated implementation is not
foreseeable in the near future.\medskip

On the other hand, we know that jet radiation is enhanced by traceable
logarithms. This makes improved predictions for QCD observables based
on resummation possible.  The general strategy is to redefine the
perturbative series from powers of $\alpha_s$ to including the
relevant logarithms; the simplified structure of these enhanced terms
then allows for a resummation to all orders.  Once the resummed form
is known we can match onto a fixed order calculation and avoid
double-counting.  For Sudakov-type logarithms a general method for
this type of resummation is available~\cite{sudakov}, and for
particular event shape observables an automated approach
exists~\cite{caesar}. In LHC analyses, the resummation of finite 
logarithms in the presence of a jet-veto scale is of 
interest~\cite{neubert,banfi_2}.\medskip

A numerical approach to resummation is provided by parton-shower 
simulations \cite{MC_review}.  It is automated in the multi-purpose 
Monte Carlo generators Pythia~\cite{pythia}, Herwig~\cite{herwig} and 
Sherpa~\cite{sherpa} to leading order in the strong coupling 
combined with the resummation of leading collinear logarithms (LO/LL).  
This method differs from the previous approaches in that the full 
spectrum of final state partons or hadrons is produced explicitly. 
While the parton shower is well defined for relatively small transverse 
momenta of the jets it is not applicable for hard jet radiation. However, 
this limitation is overcome by the CKKW~\cite{ckkw}, MLM~\cite{mlm}, 
and CKKW-L~\cite{ckkwl} jet-merging algorithms, that incorporate the
tree-level matrix-element corrections for the first few hardest 
emissions \cite{MC_review,MC_comp}. 

A complementary strategy is provided by the MC@NLO \cite{MCatNLO} and 
POWHEG \cite{powheg} approaches, that realize the matching of NLO 
calculations with parton showers. While these methods guarantee NLO/LL 
accuracy only the first/hardest shower emission gets corrected by the 
real-emission matrix element. Higher jet multiplicities are described 
in the parton-shower approximation only.  First attempts to combine 
the NLO/LL approaches with the tree-level merging ansatz have been 
reported recently~\cite{menlops}. An unprecedented level of
sophistication for predicting multi-jet final states is achieved by the 
promotion of merging algorithms to next-to-leading order accuracy 
\cite{ckkwatnlo}.\bigskip

Even though we can nowadays simulate multi-jet events, a detailed 
understanding of inclusive or exclusive $\nj$ distributions at the LHC
is still missing. Its universal features have been studied since
1985~\cite{scaling_orig}. Scaling patterns can be conveniently 
displayed in the ratio of successive exclusive jet cross-sections
\begin{equation}
R_{(n+1)/n} \; = \; \dfrac{ \sigma_{n+1} }{\sigma_n} 
= \; \dfrac{P_{n+1}}{P_n}
\qqquad \text{with}
\qquad 
P_n = \dfrac{\sigma_n}{\sigma_\text{tot}} \; .
\label{eq:def_r}
\end{equation}
We define the jet multiplicity $n$ as the number of jets {\sl in addition}
to the hard process, \eg $\sigma_1$ for pure QCD di-jets is
experimentally a 3-jet final state. Jets which are part of the hard
process are not included in the scaling analysis because they do not
arise from single QCD emissions.

Two patterns provide limiting cases for most 
LHC processes and are referred to as staircase and 
Poisson scaling.  Staircase scaling is defined as constant 
ratios between the successive multiplicity cross sections 
\begin{equation}
R_{(n+1)/n} = R \equiv e^{-b} \; ,
\label{eq:stair}
\end{equation}
where $R$ and $b$ are constant.  The exclusive $n$-jet rate for this
distribution is $\sigma_n = \sigma_0 e^{-bn}$ where $\sigma_0$ is the
0-jet exclusive cross section. Staircase scaling for exclusive and
inclusive jet rates is equivalent, with identical values of $R$.  For
a Poisson distribution with expectation value $\bar{n}$ the rates are 
\begin{equation}
P_n =  \frac{\bar{n}^n 
e^{-\bar{n}}}{n!}  \qqquad \text{or} \qqquad
R_{(n+1)/n} = \frac{\bar{n}}{n+1} \; .
\label{eq:pois}
\end{equation}
More properties of the distributions are described in
Ref.~\cite{us_prl}.\medskip

Examples of Poisson~\cite{atlas_poisson} and staircase~\cite{ex_lhc}
scaling are found in current LHC analyses, although especially with
the former, the examples are limited. Recent computations in fixed
order perturbation theory~\cite{bh_zjets_2} confirm that the staircase
pattern improves at NLO compared to the LO prediction. In this paper
we propose a number of different test-beds for both scaling behaviors,
and provide predictions for the $\nj$ ratios. In particular we
suggest two explicit measurements based on simple extensions of already 
existing analyses, namely $Z+$jets production and gap jet studies
in pure QCD events.\medskip

Establishing the origin of scaling in radiated jets from first
principles is the primary purpose of this publication.
The line of reasoning was touched on in our previous
work, but a complete treatment is lacking.  Besides constituting an
aspect of perturbative QCD on its own, a fundamental understanding of
scaling is obligatory for subsequent phenomenological applications.
Only through an underlying mechanism can we convincingly argue that
scaling patterns supplement the information provided by fixed-order 
or parton-shower calculations. \bigskip

This paper is arranged as follows.  In Sec.~\ref{sec:it_poisson} we
introduce the standard theoretical framework for the parton shower,
and elaborate on primary emissions with respect to the core process
as a necessary condition for Poisson scaling. In conjunction, we 
demonstrate in the context of generating functionals
that scaling patterns naturally emerge and can be proved to all 
multiplicities.  
In  Sec.~\ref{sec:dy_lhc} we generalize our discussion to hadron 
colliders. We introduce the corresponding generating functional 
formalism and discuss the effect of initial state PDFs, both 
through the kinematics and initial-state backward evolution. 
We close this section by showing that also in BFKL evolution the 
same scaling patterns appear as limiting cases. Having a firm 
understanding of the emergence of the scaling patterns from QCD, we can define 
cut regimes which exhibit idealized scaling. In Sec.~\ref{sec:exper} 
we propose two experimental measurements where the underlying 
hypotheses of this paper can be tested at the LHC. We conclude and 
discuss possible extensions of our ideas in Sec.~\ref{conclusions}. 

\section{Final state splittings}
\label{sec:it_poisson}

Poisson statistics describe individual events occurring repeatedly 
and independently.  For particle emissions this implies a non-trivial
assumption.  For successive photon radiation in QED it is the 
standard solution to divergences arising from soft and collinear 
photon radiation.  The underlying approximation for multiple soft photon 
radiation is the eikonal approximation, in which the radiation of a 
soft photon does not affect the hard process.
The eikonal approximation states nothing but a basic assumption of
the Poisson process --- each successive photon sees the 
same unaltered hard process. For QCD it has been known 
for some time that the non-abelian nature of soft-gluon radiation leads to 
a deviation from Poisson statistics for gluon multiplicities~\cite{poisson_qcd_orig}.
When we measure jet ratios it is important to separate general 
exponentiation~\cite{exponentiation,next_to_eikonal} from a 
Poisson distribution; to generate a Poisson distribution the 
exponentiation of soft real and virtual corrections has to include 
exactly the single-emission matrix element.

\subsection{Final-state parton shower}
\label{sec:ps}

The simulation of a LO/LL event in the parton shower approximation 
starts with the generation of a single phase space point for the partonic 
core process. The process' external (colored) lines then act as seeds 
for the subsequent parton shower evolution. Driven by unitarity they start
at the hard process scale $Q$ and finish at the shower cutoff scale 
$Q_0 \sim 1$~GeV.  Hard matrix element corrections for an 
arbitrary number of additional particles can be added to the parton 
shower using the above-mentioned matching schemes~\cite{ckkw,mlm,ckkwl}. 
However, in this section we treat all additional emissions as coming 
from the parton shower and disregard matrix element corrections.\medskip

The basis of the LL parton shower is the fully factorized form of the
collinear matrix element and phase space $d\sigma_{n+1} \sim d\sigma_n
\; P_{1\to 2} \; d\Phi_1$.  Using this simplification the parton shower remains
local, but loses information on spin, color correlations and
interference effects in addition to higher order terms neglected in the
$1\to 2$ splitting kernels $P_{1\to2}$.  Besides practicality, one of 
the benefits of collinear factorization is that the resummation of LL 
and some NLL contributions follow very naturally.  To see this, we
represent the evolution along an individual line by integrating the
Sudakov factor over the appropriate virtuality scales from the
lower cutoff $Q_0^2$ to a free hard scale $t$ 
\begin{equation}
\Delta_j(t) 
= \exp \left[ -\int_{Q_0^2}^{t} dt' \; \Gamma_j(t,t') \right] \, .
\label{eq:sudakov_def}
\end{equation}
For the regularized splitting kernels $\Gamma_j$ we use the next-to-leading
logarithmic approximation
\begin{alignat}{5}
\Gamma_j (t,t')
&\; = \;c_j \,\frac{\alpha_s(t')}{2\pi t'}
\left( \log \frac{t}{t'} - A_j \right) \; ,
\label{eq:splitting}
\end{alignat}
with color factors $c_j = C_F$ ($C_A$) and the constant terms $A_j =
3/2$ ($11/6$) for gluon emission off a quark (gluon).  The lower cutoff scale
$Q_0$ is omitted in the argument of the Sudakov form-factor. Expanding
the exponential we see that \eq{eq:sudakov_def} represents an
arbitrary number of soft and collinearly enhanced emissions, either
resolved or unresolved.\bigskip

To describe a parton-shower simulated event we note that
the QCD evolution proceeds as an integration of the product Sudakov
along the virtuality $t$,
\begin{alignat}{5}
\mathbf{\Delta}(t) 
= \prod_\text{ext lines} \Delta_j(t) 
\equiv e^{-\mathbf{\Gamma}} \; .
\label{eq:prod_sud}
\end{alignat}
The product defining $\mathbf{\Gamma}$ is over the appropriate factors
for each external line, where $j$ denotes the particle
flavor. Limiting ourselves to final state splittings this expression
only contains evolution kernels as shown in \eq{eq:sudakov_def}, and
it is by construction guaranteed to exponentiate with an appropriate
expression $\mathbf{\Gamma}$.  As long as $\mathbf{\Gamma}$ is fully
local and does not depend on previous emissions it is guaranteed to
produce a Poisson distribution for the multiplicities.  The
exponentiated form in \eq{eq:prod_sud} immediately identifies $\bar{n}
= \mathbf{\Gamma}$. This statement does not depend on the form of
$\mathbf{\Gamma}$ or its dependence on the hard scale $t$. All
that matters is that each splitting does not change the subsequent
evolution.  In the remainder of this paper we define all emissions
directly contained in the expansion of \eq{eq:prod_sud} as {\em primary}
with respect to the core process.\bigskip

\begin{figure}[t]
\centering
\includegraphics[width=0.3\textwidth]{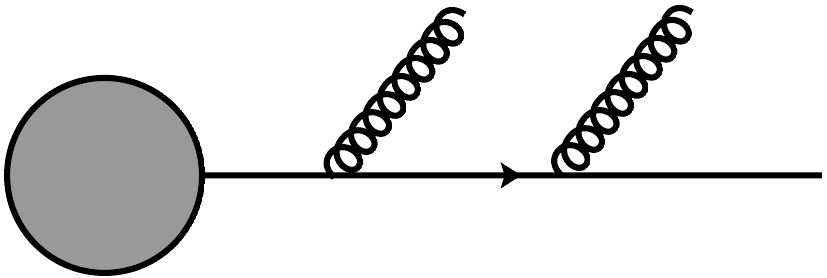}
\put(-120,12){\large$Q^2$}
\put(2,12){\large$Q_0^2$}
\hspace*{0.15\textwidth}
\includegraphics[width=0.3\textwidth]{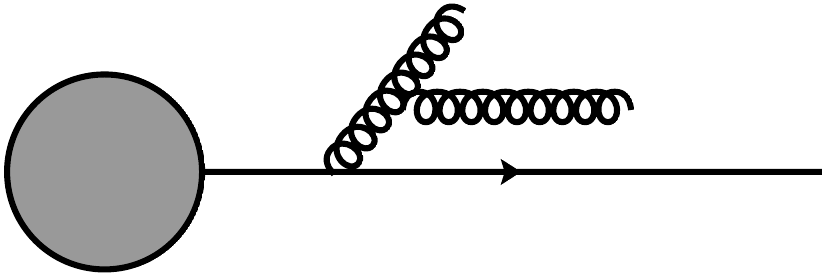}
\put(-120,12){\large$Q^2$}
\put(2,12){\large$Q_0^2$}
\caption{Simplest primary (left) and secondary
  contributions (right) assuming a core process with a hard quark
  line.}
\label{fig:abel_es}
\end{figure}

The first splitting in the parton shower picture defines the single
emission probability.  Following \fig{fig:abel_es} a second emission
can then appear from the original leg or off the first emission. For
the former, this emission is contained in \eq{eq:prod_sud} and does
not change the Poisson pattern. The latter changes the exponential; we
refer to it as {\em secondary} with respect to the original hard
process.  From a scaling perspective the relevant questions are first,
what is the relative size of the two contributions; and second if we
can change the individual strengths of primary and secondary emissions
through kinematic cuts.\medskip

In the parton shower approximation we can associate specific integrals
over virtuality with individual partonic structures appearing in
the final state evolution.  An alternative evolution ordered in a
consistent variable (\eg angle) is logarithmically equivalent.  Using
this formalism the primary contribution to two gluon emission off a
hard quark shown in \fig{fig:abel_es} is
\begin{alignat}{5}
\sigma^\text{primary}(Q^2,Q_0^2) =
 c^\text{primary} \int_{Q_0^2}^{Q^2} dt \;
\Gamma(Q^2,t) \Delta_g (t) 
 \int_{Q_0^2}^{Q^2} dt' \;
\Gamma(Q^2,t') \Delta_g (t') \; .
\label{eq:exp_c}
\end{alignat}
The coefficient $c^\text{primary}$ which includes the Sudakovs
associated with the hard line is process dependent, as this hard line
can be either a quark or a gluon.  The two external scales are the
scale $Q$ of the hard process and the lower cutoff scale $Q_0$.  If
the primary emissions are strongly ordered in the evolution variable,
the corresponding phase space factor $1/2$ is absorbed in
$c^\text{primary}$.  The simplest secondary contribution also shown in
\fig{fig:abel_es} is,
\begin{alignat}{5}
\sigma^\text{secondary}(Q^2,Q_0^2)  =
 c^\text{secondary}\int_{Q_0^{2}}^{Q^2} dt \,
\Gamma(Q^2,t) \Delta_g (t) 
 \int_{Q_0^2}^t dt' \,
\Gamma(t,t') \Delta_g (t') \; .
\label{eq:nexp_c}
\end{alignat}
The splitting kernels in the two expressions only differ in the
integral boundaries for the second emission.  In the leading
logarithmic approximation (in the exponent) for the Sudakov factors,
we can perform the integrals in \eq{eq:exp_c} and \eq{eq:nexp_c} in
terms of error functions.  The full expressions are not particularly
enlightening, but two specific limits contain crucial information.
\medskip

\begin{itemize}
\item[(1)] $\dfrac{\alpha_s}{\pi} \log^2 \dfrac{Q}{Q_0} \gg 1$\medskip

In this limit we expand \eq{eq:exp_c} and \eq{eq:nexp_c} around
$Q_0/Q \to 0$ and find the leading terms
\begin{alignat}{5}
\sigma^\text{primary} &=  \frac{ c^\text{primary}}{4}\left[ \frac{\alpha_s}{C_A} 
\log^2 \frac{Q}{Q_0} 
\;-\;\sqrt{\frac{4\alpha_s}{C_A^3}}
\log \frac{Q}{Q_0}   \;+\; \ope\left(\frac{Q_0^2}{Q^2}\right)\right]
\notag \\
\sigma^\text{secondary}& =  \frac{ c^\text{secondary}}{4}\left[   
(\sqrt{2}-1) \sqrt{\frac{\alpha_s}{C_A^3}}
\log \frac{Q}{Q_0} \;+\; \ope\left(\frac{Q_0^2}{Q^2}\right)\right] \; .
\label{eq:finsuc}
\end{alignat}
Their ratio scales like $\sigma^\text{primary}/\sigma^\text{secondary}
\propto \sqrt{\alpha_s} \, \log Q/Q_0$, \ie the primary emissions
are logarithmically enhanced.  In the limit of a large logarithm (high
single emission probability) the distribution of final state emissions
are increasingly primary, and therefore give a Poisson distribution.

Physically interpreting \eq{eq:finsuc}, a second logarithm in the
secondary contribution would come from the right-most Sudakov of
\eq{eq:nexp_c}. However, it has vanishing support for $Q\to \infty$
and does not appear in the approximate result.  The emitted gluon 
in this case spans a vanishing relative fraction in virtuality space 
where it may emit an additional parton.

\item[(2)] $\dfrac{\alpha_s}{\pi} \log^2 \dfrac{Q}{Q_0} \ll 1$\medskip

Taking this limit of \eq{eq:exp_c} and \eq{eq:nexp_c}, we find
\begin{alignat}{5}
\sigma^\text{primary}(Q^2,Q_0^2) 
&= c^\text{primary} \frac{\alpha_s^2}{4(2\pi)^2}
 \log^4 \frac{Q}{Q_0} \; + \; \ope\left(\alpha_s^3 \log^6 \frac{Q}{Q_0}
 \right) \notag \\
&= 6\, \frac{c^\text{primary}}{c^\text{secondary}} \,
\sigma^\text{secondary}(Q^2,Q_0^2) \; .
 \label{eq:smalll}
\end{alignat}
The two contributions become logarithmically equivalent and differ by
an $\ope(1)$ constant depending primarily on color factors.  In this
regime the emission probability is small and the final state is
selected democratically.  The formerly Poisson scaling pattern
receives large contributions from subsequent or secondary splittings.
Note that to justify the logarithmic expansion we still require
$\log^2 Q/Q_0 > 1$ but not large enough to spoil the small emission
probability.

\end{itemize}

In Appendix~\ref{app:toy} we use a toy model of secondary splittings
to show that $c^\text{secondary}$ gives the subsequent splitting
parameter $\bar{n}'$ in an iterated inhomogeneous Poisson
distribution. 

\subsection{Generating functional for jet fractions}
\label{sec:genfunc}

Following the argument presented in the last section we need a way to
derive scaling patterns for arbitrarily high parton multiplicities.
The generating functional formalism for QCD allows us to calculate
resummed jet quantities~\cite{basics_pqcd,esw}. We construct a
generating functional in an arbitrary parameter $u$ by demanding that
repeated differentiation at $u=0$ gives exclusive multiplicity
distributions.  Different moments of the same generating functional
then produce more inclusive jet observables.  This gives us a set of
coupled integral equations which we can solve in the limit of large
and small emission probabilities. We will find that the derived jet
multiplicity distributions follow a Poisson or staircase pattern,
respectively.\medskip

For a known series of functions $P_n$ we define 
\begin{alignat}{5}
  \Phi = \sum_{n=0}^\infty u^n P_{n-1}
\qqquad 
\text{with}
\quad 
  P_{n-1} = \left. \frac{1}{n!} \frac{d^n}{du^n} \Phi \right|_{u=0} \; .
\label{eq:def_gf}
\end{alignat}
Note that for generating functionals we always suppress the argument $u$.
In the application to gluon emission the explicit factor $1/n!$ 
corresponds to the phase space factor for identical bosons.
The exclusive jet rates $P_n$ are defined in \eq{eq:def_r}. In
accordance with that definition we only count radiated jets, for 
the generating functional that means we use $P_{n-1}$ instead 
of the usual $P_n$ consistently.\bigskip

The quark and gluon generating functionals to next-to-leading
logarithmic accuracy are
\begin{alignat}{5}
\Phi_q(Q^2) &= u \; \exp 
\left[ \int_{Q_0^2}^{Q^2} dt \; \Gamma_q(Q^2,t) 
       \left( \Phi_g(t) -1 \right) \right] 
\notag \\ 
\Phi_g(Q^2) &= u \; \exp
\left[ \int_{Q_0^2}^{Q^2} dt \left( \Gamma_g (Q^2,t)
    \left( \Phi_g(t) -1 \right) + \Gamma_f(t) \left( \frac
        {\Phi_q^2(t)} {\Phi_g(t)} -1 \right) \right) \right] \; .
\label{eq:gf_evolution}
\end{alignat}
The splitting kernels are defined in \eq{eq:splitting}; gluon
splitting to quarks, described by the kernel $\Gamma_f(t)$, is 
suppressed by a power of the logarithm. Assuming $Q \gg Q_0$, 
or a high emission probability as discussed in Section~\ref{sec:ps}, 
the largest contribution to the $t$ integration comes from the region 
where $t \approx Q_0^2$ and $\Phi_{q,g}(t) \approx \Phi_{q,g}(Q_0^2) 
\approx u$. Both evolution equations then read
\begin{alignat}{5}
\Phi_j(Q^2) = u \; \exp 
\left[ \int_{Q_0^2}^{Q^2} dt \; \Gamma_j(Q^2,t) 
       \left( u -1 \right) \right] 
= \frac{u \, \Delta_j(Q^2)}{\Delta_j(Q^2)^u} \; ,
\label{eq:gf_poisson_approx}
\end{alignat}
with the Sudakov factor defined in \eq{eq:sudakov_def}. For the jet
rates we find a Poisson distribution
\begin{alignat}{5}
P_{n-1} 
= \Delta_j(Q^2) \; \frac{|\log \Delta_j(Q^2)|^{n-1}}{(n-1)!}
\qqquad \text{or} \qquad 
R_{(n+1)/n} = \frac{|\log \Delta_j(Q^2)|}{n+1} \; .
\label{eq:gf_poisson}
\end{alignat}
Again, the jet counting reflects our convention that of $n$ jets only
$n-1$ are radiated off the hard line.  The result in
\eq{eq:gf_poisson} reflects the same underlying physics as
\eq{eq:finsuc}, namely a universal logarithmic enhancement of the
primary emission over subsequent ones. The latter also covers the sub-leading
terms to the pure Poisson distribution, and thus determines the size
of the leading corrections to \eq{eq:gf_poisson}.  For a hard quark as
well as for a hard gluon line this Poisson distribution contains only
logarithmically enhanced gluon radiation. According to the approximation 
of \eq{eq:gf_poisson_approx}, any subsequent splitting of the 
radiated gluons is subleading.\bigskip

To investigate deviations from this perfect Poisson pattern we first
study jet fractions $P_n$ for up to four emissions in $e^+ e^-$
collisions. We analytically derive them using the generating
functional, \eq{eq:def_gf}, and show the results in
Appendix~\ref{app:fractions}.  Expanding them to $\ope(\alpha_s^5)$
and combining this with inclusive unitarity $\sum P_n = 1$ at each
fixed order gives the double-logarithmically enhanced contribution for
$n \le 5$.  They show the expected Poisson pattern for the abelian
terms $\propto C_F$,
\begin{equation}
P_n = \frac{[\Delta_q(Q^2)]^2}{n!}
\left( \int_{Q_0^2}^{Q^2} dt \,\Gamma_q(Q^2,t) \Delta_g (t) \right)^n \; .
\label{eq:ll_abel}
\end{equation}
The additional gluon Sudakov compared to \eq{eq:gf_poisson} takes into
account that the radiated gluons do not split in the primary
contributions. The non-abelian terms $\propto C_A$ do not exponentiate
with respect to the $q\bar{q}$ final state. They deviate from Poisson
scaling starting at two additional jets and provide sensitivity to the
triple-gluon vertex~\cite{4j_angles}. The secondary
contribution to two-jet emission is color enhanced via $C_A / C_F$,
but ultimately smaller than the Poisson term due to the averaging
factor over the second splitting function. For the leading double
logarithm this is a simple suppression factor of $1/6$.  In
Appendix~\ref{app:toy} we compare the LL jet rates and a toy model for
an iterated Poisson process.\bigskip

In addition to purely non-abelian splittings, mixed primary and
secondary contributions also deviate from the Poisson pattern.  This
effect we can study in the average jet multiplicity in $e^+ e^-$
collisions. This observable conveniently singles out all emission
histories which are secondary with respect to the core process. This
way we generate the highest non-abelian terms at each perturbative
order,
\begin{alignat}{5}
\anj =& \; 2
       + \frac{\alpha_s}{\pi} \frac{C_F}{2}  \log^2 y_\text{cut} \\
      &+ \left( \frac{\alpha_s}{\pi} \right)^2\frac{C_A C_F}{48} \log^4 y_\text{cut} 
       + \left( \frac{\alpha_s}{\pi} \right)^3 \frac{C_{A}^2 C_F} {2880} \log^6 y_\text{cut} \notag \\
      &+ \left( \frac{\alpha_s}{\pi} \right)^4 \frac{C_{A}^3 C_F}{322560} \log^8 y_\text{cut}
       + \left( \frac{\alpha_s}{\pi} \right)^5 \frac{C_{A}^4 C_F}{58060800} \log^{10} y_\text{cut}
       + \ope \left( \left( \frac{\alpha_s}{\pi} \right)^6 \log^{12} y_\text{cut} \right) \notag \, .
\end{alignat}
The second term gives the Poisson distribution for abelian emissions
off the quark line, \ie $\bar{n}$ as given in \eq{eq:ll_abel}.  All
other terms are leading in $C_A$ for a given power of $\alpha_s$, \ie
they originate from a single radiated gluon~\cite{esw}. Mixed terms of
order $C_F^{n+1} C_A^m$ are absorbed into purely $C_F C_A^m$ terms
through exponentiation. Therefore, the combination $\anj -2 - \alpha_s
C_F/(2\pi) \log^2 y_\text{cut}$ reflects the non-Poisson nature of the
purely secondary emission at leading logarithm.\bigskip

Going back to the limiting cases introduced in Section~\ref{sec:ps} we
also find a recursive solution for small emission probabilities when 
restricting ourselves to pure Yang-Mills theory. We take $Q \sim Q_0$, 
but large enough to define logarithmically enhanced terms. The 
generating functional for a theory with only gluons satisfies
\begin{alignat}{5}
\frac{d\Phi_g(Q^2)}{dQ^2} &= 
\Phi_g(Q^2) \, \frac{C_A}{2\pi Q^2} 
 \left[ 
 - \frac{11 \alpha_s(Q^2)}{6} \left( \Phi_g(Q^2) -1 \right)
 \; + \;\int_{Q_0^2}^{Q^2} dt \frac{\alpha_s(t)}{t} \left( \Phi_g(t) -1 \right)      
 \right] \notag \\
&= 
 \; \Phi_g(Q^2) \times \label{eq:dif_eq} \\ 
&\qquad \left[ \tilde{\Gamma}_g(Q^2,Q_0^2) \,\left( \Phi_g(Q^2) -1 \right) 
       - \frac{C_A}{2\pi Q^2} \int_{Q_0^2}^{Q^2} dt \log \frac{t}{Q_0^2}  
        \left( \frac{d}{dt} \alpha_s(t) \left( \Phi_g(t) -1 \right) \right) 
        \right ] \, , \notag 
\end{alignat}
after integrating by parts and defining
\begin{equation}
\tilde{\Gamma}_g (Q^2,Q_0^2)
\; = \;C_A \,\frac{\alpha_s(Q^2)}{2\pi Q^2}
\left( \log \frac{Q^2}{Q_0^2} - \frac{11}{6} \right) \; .
\label{eq:splitting_ne}
\end{equation}
Only including the leading logarithms this is simply the negative
splitting function from \eq{eq:splitting}, but it obviously differs
beyond this approximation.  To establish the accuracy of \eq{eq:dif_eq}
including the maximal amount of NLL contributions, we expand in powers
of $\epsilon \equiv (Q^2-Q_0^2)/Q^2$,
\begin{alignat}{5}
\tilde{\Gamma}_g(Q^2,Q_0^2) \,\left( \Phi_g(Q^2) -1 \right)  
 &\; \approx \;
  \frac{C_A \alpha_s}{2\pi Q^2} \left[-\frac{11}{6} + \left(2 -\frac{11\alpha_s b_0}{3 \pi}\right)\epsilon 
\; + \; \ope (\epsilon^2) \right] \notag \\
       \frac{C_A}{2\pi Q^2} \int_{Q_0^2}^{Q^2} dt \log \frac{t}{Q_0^2}  
        \left( \frac{d}{dt} \alpha_s(t) \left( \Phi_g(t) -1 \right) \right)  
       & \; \approx \; - \frac{C_A \alpha_s}{2\pi Q^2}
        \left[ - 4 \,\epsilon^2 + \ope (\epsilon^3) \right] \, .
\end{alignat}
The logarithmic and finite contribution from the first term on the RHS
of the differential equation are of order $\epsilon^1$ and
$\epsilon^0$ while the formally infinite series from the second term
in \eq{eq:dif_eq} starts at $\epsilon^2$. Keeping only terms linear in
$\epsilon$ we obtain the simple form
\begin{alignat}{5}
 \frac{d\Phi_g(Q^2)}{dQ^2} & 
\,\approx \,\Phi_g(Q^2) 
  \; \tilde{\Gamma}_g(Q^2,Q_0^2) \left( \Phi_g(Q^2) -1 \right) \; .
\end{alignat}  
Including the boundary condition $\Phi_g(Q_0^2) = u$ we can solve this,
\begin{alignat}{5}
\Phi_g(Q^2) 
&= \dfrac{1}{1 + \dfrac{(1-u)}{u \tilde{\Delta}_g(Q^2)}} 
\qqquad \text{with} \qquad 
\tilde{\Delta}_g(Q^2) = \exp \left[ - \int_{Q_0^2}^{Q^2} dt \tilde{\Gamma}_g(t,Q_0^2) \right] \; .
\label{eq:gf_solution}
\end{alignat}
Neglecting the effects of the running coupling, $\tilde{\Delta}_g(Q^2)$
is a Sudakov form factor. Including the running coupling,
\eq{eq:gf_solution} differs from the standard Sudakov in
\eq{eq:sudakov_def} starting at higher orders,
\begin{equation}
\frac{\tilde{\Delta}_g(Q^2)}{\Delta_g(Q^2)} \;  =\; 
\exp \left( - \frac{\alpha_s^2}{12 \pi} b_0 \, 
\log^3 \frac{Q^2}{Q_0^2} \right) \; .
\label{eq:sudakov_dif}
\end{equation}
Taking derivatives of the generating functional in \eq{eq:gf_solution} 
at $u=0$ we can compute the exclusive jet rates
\begin{alignat}{5}
P_{n-1} = \tilde{\Delta}_g(Q^2) \left( 1 - \tilde{\Delta}_g(Q^2) \right)^{n-1} 
\qqquad \text{or} \qquad 
R_{(n+1)/n} = 1 - \tilde{\Delta}_g(Q^2) \; .
\label{eq:gf_staircase}
\end{alignat}
These constant ratios define a staircase pattern.  Comparing
\eq{eq:gf_poisson} and \eq{eq:gf_staircase} we see that in two
distinct phase space regimes we find two clear scaling patterns for
the Yang-Mills or pure gluon case. Both of them can arise in final
state gluon radiation, which means they should in principle be
observable in $e^+ e^- \to$ jets events.\bigskip

\begin{figure}[t]
\centering
\includegraphics[height=0.49\textwidth]{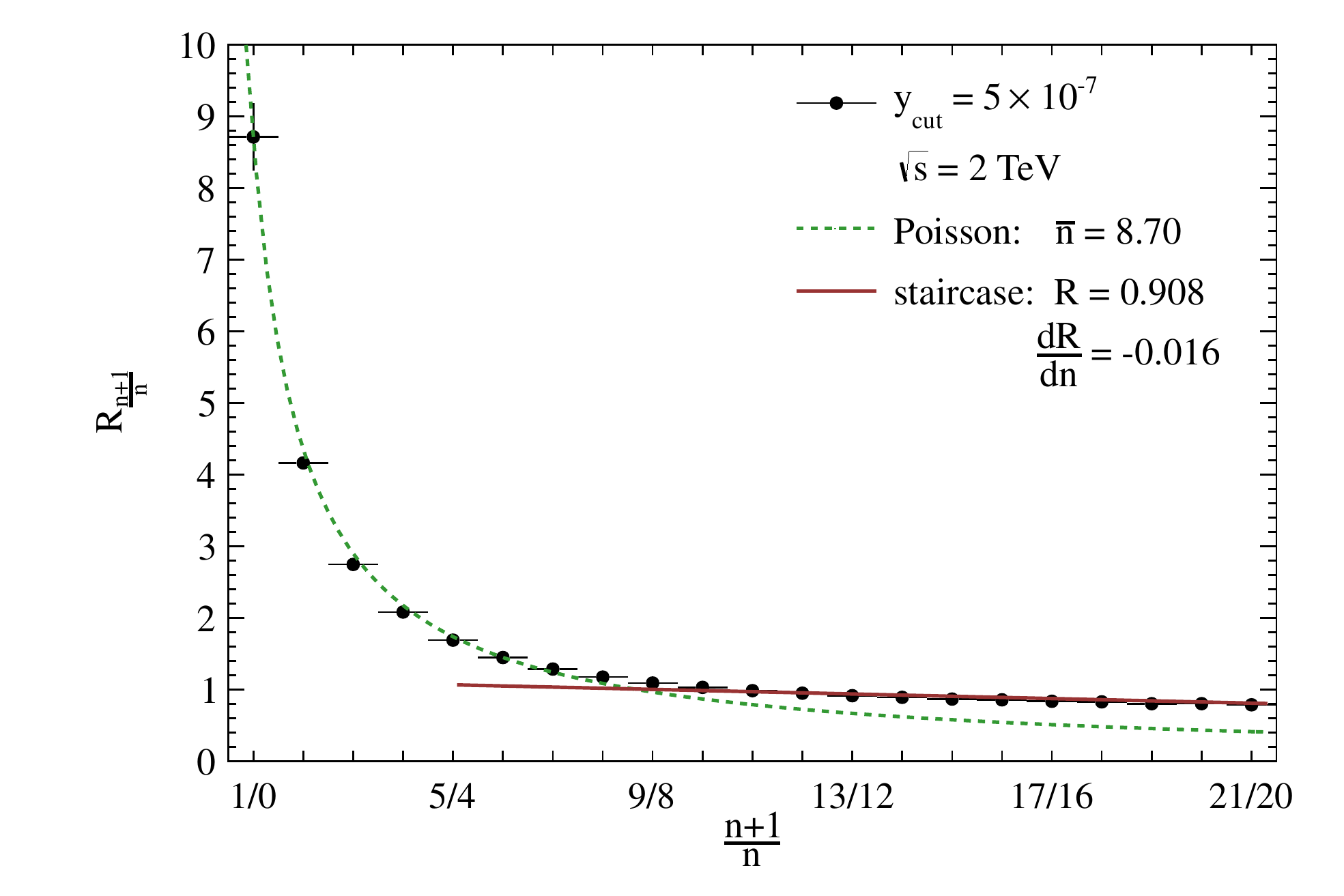}
\caption{Jet ratios $R_{(n+1)/n}$ in $e^+ e^- \to$~jets production at
  2~TeV center-of-mass energy. We show a Poisson fit with
  $\bar{n}=8.7$ and a staircase fit to the tail. We use
  Sherpa~\cite{sherpa,sh_shower} with the $g\rightarrow q\bar{q}$
  shower splittings switched off.}
\label{fig:ee_staircase}
\end{figure}

The all-order theoretical predictions for Poisson scaling,
\eq{eq:gf_poisson}, and staircase scaling, \eq{eq:gf_staircase}, we
can compare to simulated $e^+ e^- \to$~jets events.  To cover both, a
large scale separation $Q \gg Q_0$ as well as a democratic scale $Q
\sim Q_0$, we use a large center-of-mass energy of 2~TeV and a very
small lower cutoff $y_\text{cut} = 5\cdot 10^{-7}$ for the Durham
jet-reconstruction algorithm~\cite{new_clust}.  In
\fig{fig:ee_staircase} we show jet ratios $R_{(n+1)/n}$ for a large
range of $n$. Indeed, we observe Poisson as well as staircase
scaling. The same behavior is known from hadron colliders for example
in $pp \to \gamma$+jets production~\cite{us_photon}: for relatively
low $n$ values the emission is dominated by large scale differences,
inducing a Poisson pattern. For large jet multiplicity individual
emissions are not affected by a large scale difference, so we see a
staircase tail. While this transition is a solid QCD prediction it has
not been studied experimentally (yet).

\subsection{Matrix element corrections}

In all of the above discussion we only assume logarithmically induced
emission and neglect any kind of phase space effects.  A simple
test case for the relative contributions of primary vs subsequent
emissions including additional phase space information is two-gluon
emission from a $q\bar{q}$ dipole.  The squared matrix
element for strongly ordered two-gluon emissions is~\cite{att_fun}
\begin{equation}
|\mathcal{M}(p_1,p_2)|^2 = \frac{32 C_F}{p_{T,1} p_{T,2}}
\left[ C_A \left( \frac{\cosh (\eta_1 - \eta_2)} { \cosh (\eta_1 - \eta_2) 
- \cos(\phi_1 - \phi_2) } -1\right) + 2 \, C_F \right] \; ,
\label{eq:qqgg}
\end{equation}
where $\eta_i$ are the gluon rapidities, $\phi_i$ are the azimuthal
angles and $p_{T,i}$ the transverse momenta.  The term
proportional to $C_F$ with its very simple kinematic structure
represents the primary emissions.  The $C_A$ term 
corresponds to the subsequent emission contribution and contains 
an interesting dependence on the phase space.\medskip  

The transverse momentum integration leads to $\log Q/Q_0$ terms for
both contributions, which we now assume to not be too large --- otherwise 
we would be logarithmically dominated and the results from the previous 
section would apply. Instead, we are interested in the $\eta$ and $\phi$ 
dependence.  The question is whether such final state kinematics can be 
observed as deviations from our scaling patterns.  

\begin{figure}[t]
\centering
\includegraphics[height=0.49\textwidth]{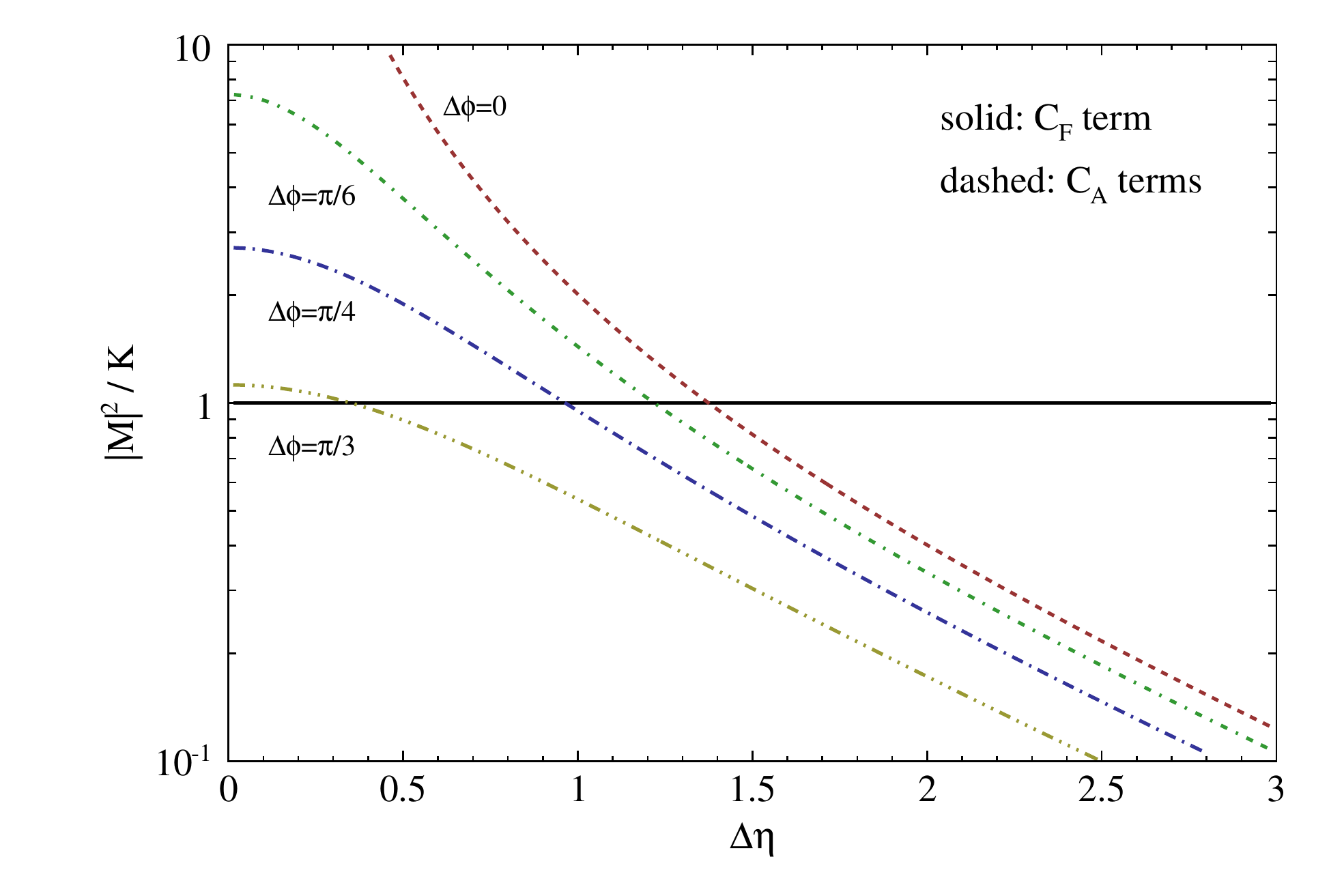}
\caption{Contribution from secondary emission to the squared matrix
  elements given in \eq{eq:qqgg} as a function of $\Delta \eta$ for
  $\Delta \phi = 0,\pi/6, \pi/4, \pi/3$ versus the constant primary
  contribution (horizontal line). All curves are normalized to $K = 64
  C_F^2/(p_{T,1} p_{T,2})$.}
\label{fig:qqgg}
\end{figure}

In Fig.~\ref{fig:qqgg} we examine the primary and subsequent
contribution as a function of $\Delta \eta$. Testing different $\Delta
\phi$ values, we see that the secondary emission dominates when the
gluons are close.  For $\Delta \phi = \pi/2$ the $C_A$ term vanishes
for all rapidity separations.  Note that for $\sqrt{(\Delta
  \eta)^2+(\Delta\phi)^2 } \le R$ the two emissions are clustered in a
single jet of radius $R$ (typically $R\approx 0.4 - 0.7$) and as such
do not contribute to the 4-jet rate. This suggests that for large
angle emissions the $\nj$ spectrum will remain Poisson even including
matrix element information.  Or in other words subsequent emissions
breaking the Poisson pattern are unlikely to be widely
separated.\bigskip
 
Although \eq{eq:qqgg} contains particular matrix element information,
it is still based on the eikonal approximation and does not include
kinematical constraints.  However, we know that subsequent emissions
lead to deviations from Poisson scaling in all phase space regions.
In order to get a handle exclusively on effects from outside the soft
or soft-collinear regime, we consider the photon multiplicities in the
QED process $e^+ e^- \to e^+ e^- + n\gamma$. To leading order all
photon emissions are primary.

\begin{figure}[t]
\centering
\includegraphics[height=0.49\textwidth]{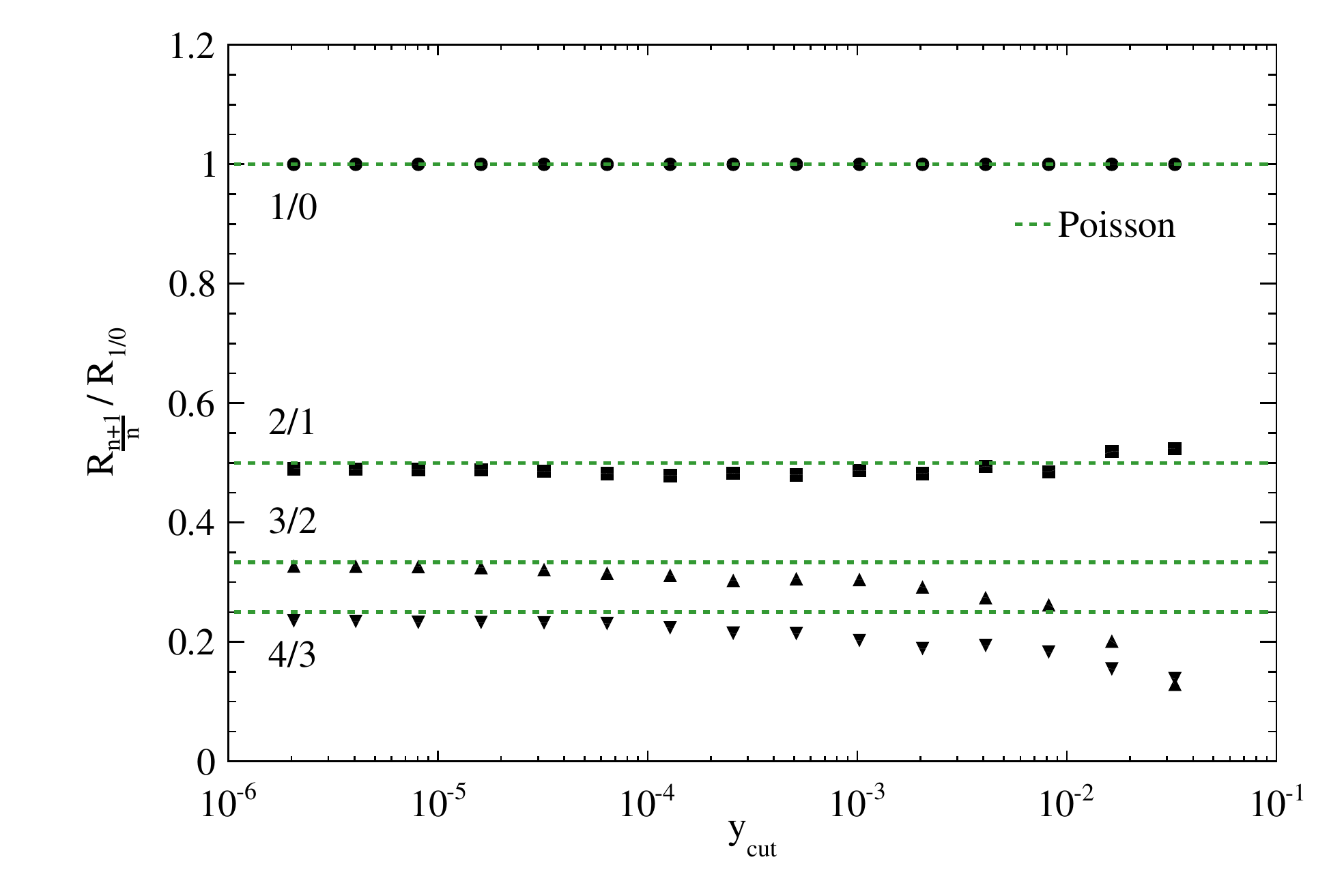}
\caption{Normalized ratios for the photon multiplicity in $e^+ e^- \to
  e^+ e^- + n\gamma$ as a function of the resolution parameter
  $y_\text{cut}$.  The solid lines correspond to the perfect Poisson
  hypothesis. We use Sherpa~\cite{sherpa}.}
\label{fig:ee_fini}
\end{figure}

We consider the cross sections for the processes 
$e^+ e^- \to e^+ e^- + n\gamma$ at $\sqrt{s}=500$ GeV using the exact 
tree-level matrix elements regulated using the $k_T$ measure
\begin{equation}
\frac{2\min(E_i^2,E_j^2)}{s} \; (1 - \cos \theta_{ij}) \; > \; y_\text{cut} \; .
\end{equation} 
For small values of $y_\text{cut}$ we should find a Poisson pattern in
the exclusive photon rates, which we confirm in \fig{fig:ee_fini}. For
larger $y_\text{cut}$ the different multiplicity distributions start
deviating from the Poisson pattern. The ratios are pushed apart from
one another, opposite to what we expect from a staircase pattern. The
reason is that each emission takes a non-negligible amount of the
total energy of the event and suppresses the phase space for
subsequent emissions. Going back to the two main scaling patterns this
means that matrix element and final-state phase space effects are not 
responsible for the transition from Poisson to staircase scaling.

\section{Hadron colliders}
\label{sec:dy_lhc}

The analytic form of the fixed energy jet fractions given in
Appendix~\ref{app:fractions} indicates that a final state cascade
initiated by a $q\bar{q}$ pair follows neither a Poisson nor a
staircase scaling pattern at low multiplicities.  However, from
$Z$+jets production we know that essentially all jet ratios are
constant, with an even stronger suppression of
$R_{1/0}$~\cite{atlas_zjets}.  This suggests that additional effects
drive the jet ratios at hadron colliders towards a staircase
pattern. One possible cause is that incoming partons do not on average
carry the same energy fractions $x$ for different final state jet
multiplicities; in that case we might observe an initial-state phase
space effect. Second, jets at hadron colliders are typically generated
through initial state radiation, and we know that the initial-state
parton shower behaves somewhat differently from final state
splittings.

\subsection{Generating functional for incoming hadrons} 
\label{sec:hadron_gf}

The basis of the QCD treatment of hadron collider physics is collinear
factorization which allows us to employ the generating functional
method~\cite{esw,basics_pqcd,had_gf}. Before we can apply any of
this to jet counting we need to clarify our choice of the
factorization scale $\mu_F$ in exclusive $n_\text{jet}$ rates, \ie in
the presence of a jet-counting or jet-veto scale $p_V$.  The
resummation properties of the DGLAP equation identify the combined
renormalization and factorization scale with a collinear cutoff below
which initial state splittings are unresolved and influence only the
functional dependence on the partonic energy fraction $x$. Because we
are interested in radiated jets with $p_T \ge p_V$ we identify the
factorization and the jet-veto scale, \ie $\mu_F \equiv p_V$. Note
that this choice furthermore avoids generating additional finite
though potentially large logarithms in the ratio
$\mu_F/p_V$~\cite{neubert}.\medskip

Symbolically, going from final state radiation in $e^+e^-$ collisions
to deep inelastic scattering (DIS) with initial state radiation and
parton densities we replace the two generating functionals,
distinguishing time-like from space-like splittings,
\begin{alignat}{5}
\Phi_q (Q^2,p_V^2) \times \Phi_{\bar{q}} (Q^2,p_V^2) 
\quad \rightarrow \quad
\Phi_{q/\bar{q}} (Q^2,p_V^2) \times \mathcal{Z}_{q/\bar{q}}(x,Q^2,p_V^2) .
\label{eq:symb1}
\end{alignat}
As in Sec.~\ref{sec:genfunc} we omit the argument $u$ in all
generating functionals.  In the original DIS context all scales are
defined in terms of the $e^+e^-$ Durham algorithm~\cite{had_gf},
most notably the hard scale $Q$ and $p_V \equiv \mu_F$ as well as the
softer resolution scale $Q_0 \leq \mu_F$. We identify all three
relevant scales $Q_0 = \mu_F = p_V$. For the DIS analysis this
corresponds to not further resolving the original macro-jets which
define the separation of resolved jets and beam jets~\cite{had_gf}.
Again, this choice omits potentially large finite scale logarithms in
our perturbative treatment.\bigskip

We also introduce an explicit $x$ dependence in the generating
functional for incoming partons as it is clear that PDF effects alter
the possibility to radiate jets.
Each emission takes away an energy fraction $1-z$ of the emitter; the
$x$ value has to change correspondingly and splitting between
different partons needs to be taken into account.  From the
factorization theorem we know that PDFs and partonic cross-sections
also factorize at the generating functional level,
\begin{alignat}{5}
   \mathcal{Z}_a(x,Q^2,p_V^2) &= \sum_b \int \limits^1_x
   \frac{dz}{z} ~f_b \left( \frac{x}{z},p_V^2 \right) \mathcal{Z}_a^b
   (z,Q^2,p_V^2) \; .
\label{eq:pdf_gf}
\end{alignat}
The parton densities we consistently evaluate at the scale $p_V$.
This way, logarithmically enhanced parton splittings above $p_V$ are
described by the partonic generating functional $\mathcal{Z}_a^b$.
For the generating functional in DIS we start with a time-like
generating functional for a single (anti-)quark and weight it with the
proper electromagnetic coupling~\cite{had_gf}
\begin{alignat}{5}
  \Phi_\text{DIS} &= \sum_a e_a^2 ~ \Phi_a (Q^2,p_V^2) ~
  \mathcal{Z}_a (x,Q^2,p_V^2) .
\end{alignat}
The partonic cross-sections and jet evolution are the same for the
quark and the anti-quark, but the PDFs are different. In DIS the 
final state kinematics fix $x$.  Additional jets radiated off the 
incoming parton imply that in this case we probe higher $x$ values 
as given by the convolution in \eq{eq:pdf_gf}.\medskip

The task is to find the evolution equations for $\mathcal{Z}_a^b$. To
leading logarithm (LL) this turns out to be relatively simple.  In the
soft and collinear limit~\cite{esw,basics_pqcd} the eikonal
approximation implies $z\approx 1$. Furthermore, the $g \rightarrow
q\bar{q}$ splitting is logarithmically suppressed compared to the
other splittings, so we can neglect it. Under these two LL assumptions 
the evolution equation and the corresponding generating function in
\eq{eq:pdf_gf} read~\cite{had_gf}
\begin{alignat}{5}
\mathcal{Z}_a^b(z,Q^2,p_V^2)  &= 
\delta(1-z) \; \delta_a^b \; \Phi_a(Q^2,p_V^2)  \notag \\
  \mathcal{Z}_a(x,Q^2,p_V^2) &= f_a(x,p_V^2)~
  \Phi_a(Q^2,p_V^2) \; .
\label{eq:hadcol_kin}
\end{alignat}
The PDF effects and the jet generating function factorize in $x$, so
we can treat them independently. In general, $\Phi_a$ is a two-scale
generating functional~\cite{had_gf,twoscale}. Because we identify $Q_0
= p_V$ the second scale is suppressed and its evolution equation is
almost the same as in the $e^+e^-$ case. To leading logarithm we
find
\begin{alignat}{5}
\Phi_a (Q^2,p_V^2) &= \exp
\left[ \, \int_{p_V^2}^{Q^2} dt \; \Gamma_a(Q^2,t) \;
          \Big( \Phi_g(t,p_V^2) -1 \Big) 
\right] \; .
\label{eq:twoscale_gf}
\end{alignat}
Compared to \eq{eq:gf_evolution} the factor $u$ in front of the
exponential is missing. The reason is that we cannot resolve a jet if
there is not at least one space like splitting. The hard parton cannot
produce a final state jet, so we always find the normalization
condition $\Phi_a(p_V^2,p_V^2) \equiv 1 $. The further evolution of
emitted partons we describe with the time-like functional of
\eq{eq:gf_evolution}.\bigskip

Moving on to Drell-Yan production with two incoming partons we need to
replace the generating functionals, symbolically written, to
\begin{alignat}{5}
\mathcal{Z}_{q/\bar{q}}(x_a,Q^2,p_V^2) 
\times 
\mathcal{Z}_{\bar{q}/q}(x_b,Q^2,p_V^2) \; .
\end{alignat}
Thus, we replace the remaining time-like generating functional with a
space-like generating functional to describe two incoming partons. A
major complication is that the final state phase space does not fix
$x_{a,b}$ anymore. Instead, we have to integrate over their allowed
ranges and find the generating functional for the Drell-Yan process,
\begin{alignat}{5}
\Phi_\text{Drell-Yan} &= 
\sum_{a,b} \int dx_adx_b~ \mathcal{Z}_a (x_a,Q^2,p_V^2) 
                \;  \mathcal{Z}_b (x_b,Q^2,p_V^2) \notag\\ &\approx
\sum_{a,b} \int dx_adx_b~ f_a(x_a,p_V^2) f_b(x_b,p_V^2) 
          \; \Phi_a (Q^2,p_V^2) \Phi_b (Q^2,p_V^2) \notag \\ &\approx
\sum_{a,b} f_a( x^{(0)}, p_V^2) \Phi_a (Q^2,p_V^2) \; 
          f_b( x^{(0)}, p_V^2) \Phi_b (Q^2,p_V^2) \; .
\label{eq:drellyan_gf}
\end{alignat}
From this generating functional we can derive the individual $n$-jet
rates.  For the second line of \eq{eq:drellyan_gf} we use the leading
logarithmic approximation as in the DIS case. To arrive at the third
line we replace the variable $x_{a,b}$ values by a typical partonic
energy scale $x^{(0)}$. For typical hadron colliders processes we
assume this value to be close to threshold and equal for the two
incoming partons. The argument $u$ which generates the different
$n$-jet rates is carried only by the generating functionals
$\Phi_{a,b}(Q^2,p_V^2)$.  Starting with two generating functionals for
the two initial state particles, hard jet radiation with $p_T > p_V$
indeed factorizes from a PDF factor.\bigskip

One apparent contradiction related to the PDF kinematics we need to
resolve.  On the one hand, in \eq{eq:hadcol_kin} the eikonal
approximation allows us to set $z\approx 1$, which means that the
entire energy dependence is encoded in the PDF factor. On the other
hand, each resolved jet requires a finite $p_T > p_V$.  Hence, the
integration range for $x_{a,b}$ is determined by the partonic $n$-jet
process and $x^{(0)}$ implicitly depends on $u$. This implicit
dependence we have to account for by hand. In particular for parton density
regimes which increase towards small $x$ the majority of multi-jet
events at the LHC are produced at threshold. The threshold value for
any of the $n$-jet production rates we denote as $x^{(n)}$, leading us
to the modified factorized form
\begin{alignat}{5}
  \Phi_\text{Drell-Yan} &= \sum_{a,b} 
  f_a( x^{(n)} , p_V^2) \Phi_a (Q^2,p_V^2) \; 
  f_b( x^{(n)}, p_V^2 ) \Phi_b (Q^2,p_V^2) \; .
  \label{toughq}
\end{alignat}
We emphasize that $n$ is determined \emph{a posteriori} upon
differentiation with respect to $u$, so is presented for illustrative
purposes only. \eq{toughq} means that to leading logarithm the jet
radiation pattern in the Drell-Yan case is the same as in $e^+e^- \to$
jets processes, modulo explicit PDF factors estimated using an
$n$-dependent threshold kinematics.  A similar approach can account
for energy momentum conservation in soft-gluon
resummation~\cite{jet_veto_simo}. This way we leave the LL evolution
of jets untouched and instead shift the $x$ value in the PDFs to
account for additional jets.  All our findings from
Sec.~\ref{sec:it_poisson} we can immediately apply, once we understand
the PDF correction factor in the next section.

\subsection{Parton density suppression}
\label{sec:hadron_pdf}

In Sec.~\ref{sec:hadron_gf} we have learned that to leading
logarithmic accuracy the effects of the parton densities and jet
emission factorize. For large jet multiplicities this explains the
observed staircase scaling at hadron 
colliders~\cite{scaling_orig,ex_lhc}.  Parton densities contribute 
to this effect in particular at low multiplicities. When increasing 
the jet multiplicity the typical partonic energy fractions $x$ 
probed by the partonic process increase as well. The relative increase 
in $x$ is largest for low jet multiplicities.

In terms of the assumed threshold kinematics adding a jet with finite
transverse momentum implies $x^{(n+1)} > x^{(n)}$.  To compute the
relative cost of producing an additional jet we estimate the ratio of
PDF values evaluated at $x^{(n)}$ and $x^{(n+1)}$ as a function of the
number of extra jets $n$.  In effect this is the discretized second
derivative with respect to $x$. For hadron collider processes
involving two parton densities $f(x,Q)$ we define the PDF correction
factor to the ratio of successive jet ratios $R_{(n+1)/n}/R_{(n+2)/(n+1)}$
\begin{alignat}{5}
B_n \; = \; 
\left|
\dfrac{ \quad \dfrac{f(x^{(n+1)}, Q)}{f(x^{(n)}, Q)} \quad }
      { \quad \dfrac{f(x^{(n+2)}, Q)}{f(x^{(n+1)},Q)} \quad }
            \right|^2 \; .
\label{pdf_dis}
\end{alignat}
The square in the definition of $B_n$ reflects the two PDFs in 
hadron collisions.  If for example the partonic ratio of two
successive jet ratios is $R_{(n+1)/n}/R_{(n+2)/(n+1)} \sim c$ then the
proper hadronic ratio becomes $B_n c$. We fix $Q$ for simplicity, but
this only mildly affects our results.

The main effects are, first, that $B_n < 1$ in most cases. This way
PDF effects suppress the lower multiplicity ratios $R_{(n+1)/n}$.  For
large jet multiplicities the relative impact of yet another jet
becomes small, $B_n \to 1$. The hadronic initial-state effect on the
jet scaling disappears and we are back to the staircase pattern.
Second, the PDF effect is largest for the steep gluon densities, as
compared to the flatter valence quarks.  Finally, allowing for
variable $Q$ the PDF values $f(x,Q)$ increase (decrease) with $Q$ for
low (high) $x$, with a cross-over point around $x \sim 0.1$.  For
small $x$ values the initial state evolution then suppresses jet
ratios at high multiplicity or large $Q^2$.\medskip

\begin{figure}[t]
\centering
\includegraphics[width=0.495\textwidth]{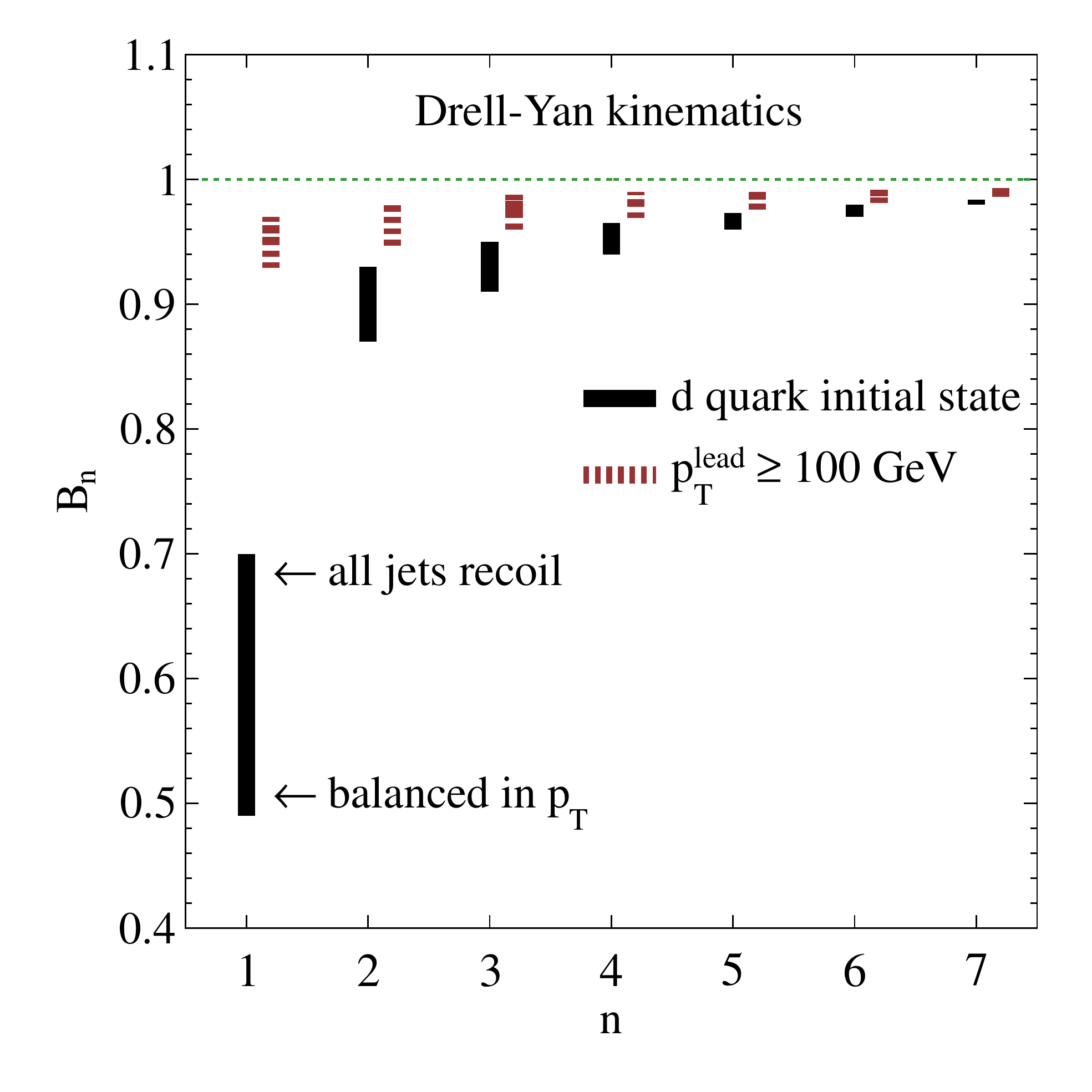}
\hfill
\includegraphics[width=0.495\textwidth]{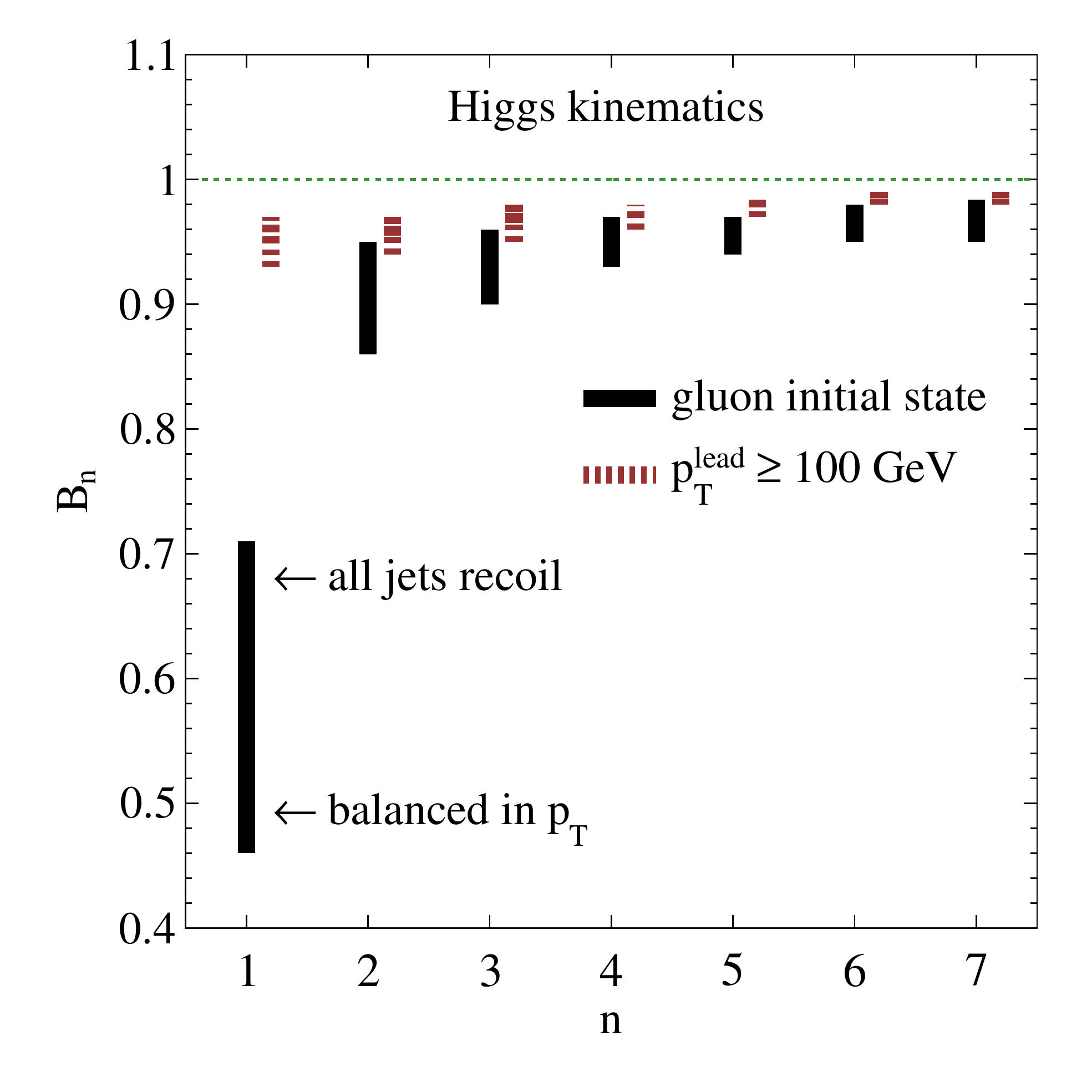}
\caption{Left panel: estimated PDF suppression for inclusive (solid)
  and jet-associated (dashed, $p^\text{lead}_T \ge 100$~GeV) Drell-Yan
  kinematics.  We assume an initial state with $d$-quarks only. Right
  panel: same for Higgs production in gluon fusion with $m_H=
  125$~GeV.  The uncertainty encompasses two representative 
  kinematical limits of the multi-jet final state, described in 
  the text.}
\label{fig:pdf_dy}
\end{figure}

What we are most interested in are PDF effects for the Drell-Yan
process at lower multiplicities. We consider the threshold values
$x^{(n)}$, for example for producing an on-shell $Z$-boson and one
additional jet,
\begin{equation}
x^{(1)} = 
\frac{\sqrt{m_Z^2 + 2 \, (p_T \sqrt{p_T^{2} + m_Z^2} + p_T^2)}}
     {2\, E_\text{beam}}
 \; .
\end{equation} 
where $E_\text{beam} = 3500$~GeV for the LHC in 2011.  
Comparing $x^{(1)}$ with $x^{(0)} \approx m_Z/(2
E_\text{beam})$ shows a sizeable shift.  For the two-jet threshold 
$x^{(2)}$ two limiting cases are either the additional jet adding 
merely $p_T/(2 E_\text{beam})$ to $x^{(1)}$ or two approximately 
collinear jets recoiling against a hard $Z$. The variation between 
these two cases estimates the uncertainty on our method which can 
be generalized straightforwardly to the $n$-jet final state. 

In the left panel of \fig{fig:pdf_dy} we display $B_n$ for the
estimated Drell-Yan kinematics, assuming each jet has transverse 
momentum $p_T=p_V=30$~GeV.  The effect on the first jet ratios is
large, but quickly diminishes towards higher $n$.  We also see that if
we require a leading jet with large transverse momentum, 
$p^\text{lead}_T \ge 100$ GeV, we move to sufficiently
high $x$ such that additional jet ratios are unaffected by the PDF
effect.  It is reassuring to see that if we combine the PDF
suppression of $R_{1/0}$ (0.46 - 0.65) with the $C_A$ enhancement of
$R_{2/1}$ (1.36) and assume an original Poisson scaling we find
$R_{1/0}/R_{2/1} = (0.67-0.95)$, in nice agreement with ATLAS
data~\cite{atlas_zjets}. This beautifully illustrates that staircase
scaling at large multiplicities can be derived from first principles
QCD while for small multiplicities it is something like a sweet spot.

As an additional check, we present the PDF suppression in gluon-fusion
Higgs production in the right panel of \fig{fig:pdf_dy}. We assume
$m_H = 125$~GeV, ignore flavor changes and consider jets with 
$p_T =p_V =30$~GeV.  The gluon PDF drops more rapidly for increasing $x$,
inducing a large PDF suppression. On the other hand, the increasing
energy of the core process as compared to the Drell-Yan process
slightly decreases the effect.  The combination of the two gives
remarkably similar results to the Drell-Yan process.

\subsection{Initial-state parton shower}
\label{sec:hadron_shower}

As indicated above, jet radiation at hadron colliders is generated
mostly through initial state radiation, which means that our
final-state analysis of Sec.~\ref{sec:it_poisson} should be modified.
We need to compute the spectrum of jets arising as primary emission in
the backward evolution and acting as seeds for subsequent final-state 
radiation. For simplicity we just consider the backward evolution 
along an initial-state quark line with a single type of branching, namely 
gluon emission.  The evolution proceeds through the space-like Sudakov 
form factor 
\begin{equation}
\Pi(t_1,t_2; x) =\exp \left \{-\int_{t_1}^{t_2} \frac{dt}{t} \int \, \frac{dz}{z} 
\,\frac{\alpha_s}{2\pi}\, P_{q\to qg}(z) \,\frac{f_q(x/z,t)}{f_q(x,t) } \right\} \; ,
\label{eq:pyb}
\end{equation}
cf. Ref.~\cite{be_tor}, with the appropriate splitting kernels for 
gluon emission $P_{q\to qg}$ and ignoring potential initial-state flavor 
changes here. The evolution of each initial state parton starts with 
momentum fraction $x_i$, determined by the hard process, and virtuality 
$t = x_1 x_2 S$, it terminates at the hadronic scales $x \approx 1$ and 
$t = Q_0^2$ associated with the transition to non-perturbative 
physics.\medskip

We know from Sec.~\ref{sec:it_poisson} that a single time-like Sudakov form
factor produces perfect Poisson scaling. Indeed, whenever we
have a non-emission probability represented by $e^{-\mathbf{\Gamma}}$ as in
\eq{eq:prod_sud}, where $\mathbf{\Gamma}$ does not change as a result of a
splitting, the process is guaranteed to produce Poisson scaling. For 
backwards evolution the situation, however, is different. Once a splitting 
is generated, e.g. using a veto algorithm~\cite{sud_veto}, we need to 
re-compute $x \to x/z$ because each emitted parton increases 
the combined $x$ value. The evolution then proceeds with this different 
effective splitting kernel.  In other words, the PDF dependence in 
\eq{eq:pyb} explicitly correlates parton emissions, breaking a key 
ingredient to the Poisson derivation.\medskip

\begin{figure}[t]
\centering
\includegraphics[height=0.49\textwidth]{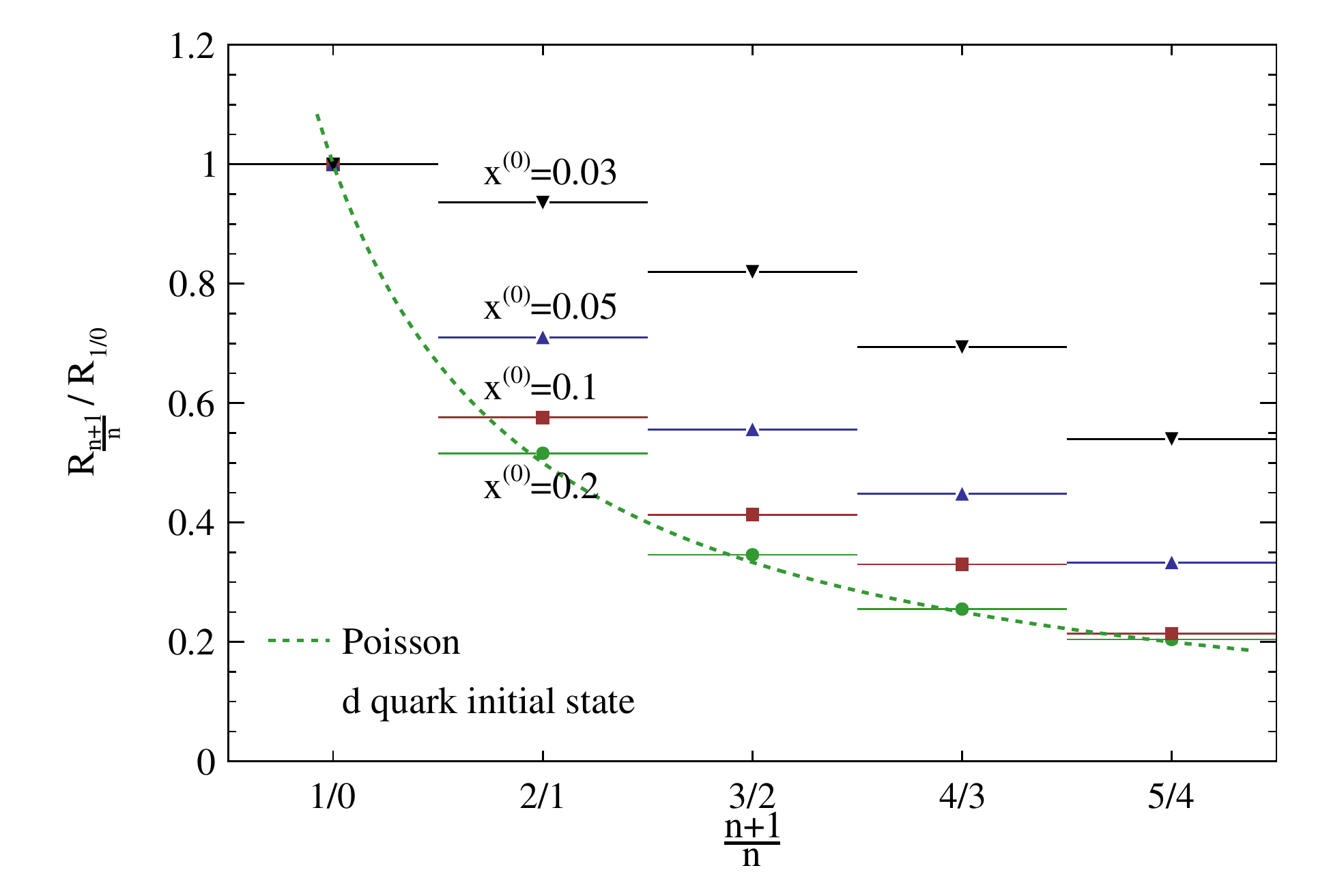}
\caption{Normalized ratios of gluon production rates originating
from the backward evolution of $d$-quarks according to \eq{eq:pyb}.
We assume different initial values for $x^{(0)}$.  Each splitting 
is restricted to $z$ values generating a minimal increase of 
$\Delta x = 0.02$. The dashed line indicates the expectation
for a perfect Poisson distribution of the underlying jet rates.}
\label{fig:bw}
\end{figure}

To quantify this effect we numerically evaluate the gluon 
emission spectrum generated by the Sudakov form factor given in 
\eq{eq:pyb} using a veto algorithm. We neglect any recoil effects 
and therefore expect a somewhat smaller suppression in the first bin. 
In \fig{fig:bw} we display the normalized ratios of the gluon production 
rates off initial state $d$-quarks. We assume different starting values 
for $x^{(0)}$ thereby keeping the evolution distance and the minimum 
step size of $\Delta x = 0.02$ fixed. This roughly corresponds to the 
emission of a hard additional jet with $p_T \approx 70$~GeV which we 
are interested in for the LHC. The starting scale we vary between 
$x^{(0)} = 0.03$ and $x^{(0)} = 0.2 $ corresponding to a hard process 
of $105$~GeV and $700$~GeV respectively. The jet ratios strongly deviate 
from the Poisson pattern for low $x^{(0)}$, while the effect quickly 
diminishes for large $x^{(0)}$.  This is the same pattern we 
find for the PDF effect in Sec.~\ref{sec:hadron_pdf}.
\bigskip

Finally, to see how our two approaches to parton density effects are
related we study the PDF part of the weight attached to a single
resolvable emission coming from a collision with momentum fraction
$x_0$. Using the backward evolution \eq{eq:pyb} the exclusive one-jet
rate can be represented as
\begin{alignat}{5}
\sigma_1 \; \sim \; &
f_1(x^{(0)},Q^2)\, f_2(x^{(0)},Q^2) \; \sigma^{\text{partonic}}_0 \;
  \int  \frac{dt}{t}  \int \frac{dz}{z} 
  \, P(z) \frac{f(x^{(0)}/z , t)}{f(x^{(0)},t)} 
\notag \\
\sim \; &
f_1(x^{(0)},Q^2)\, f_2(x^{(0)},Q^2) \; \sigma^{\text{partonic}}_0 \;
\frac{f(x^{(1)},Q^2)}{f(x^{(0)},Q^2)} \;
  \, \frac{P(z^{(1)})}{z^{(1)}} \; . 
\label{eq:pdf_just}
\end{alignat} 
In the second line we limit ourselves to the leading logarithmic
approximation (\ie ignoring the $t$ dependence of the PDFs) and fix
the resolvable momentum fraction to its threshold value $z^{(1)}=x^{(0)}/x^{(1)}$. 
The effect of the PDF weight in the ratio $\sigma_1 / \sigma_0$ then 
turns into a suppression factor $f(x^{(1)}, Q^2)/ f(x^{(0)}, Q^2)$.  Each
emission just pushes up the overall PDF suppression and we are
effectively led to the estimate on the shape of the ratios provided by
\eq{pdf_dis}.

\subsection{BFKL evolution}
\label{sec:hadron_bfkl}

So far, we only consider the DGLAP evolution which relies on
collinear factorization and resums collinear logarithms.
Parton evolution can also be represented by BFKL~\cite{bfkl} or
CCFM~\cite{ccfm} dynamics which rely on an entirely different form of
the factorized matrix element.  In this approach a simple expression
for the $n$-jet generating function at leading logarithmic order in
$\log 1/x$ and $\log Q/p_V$ reads~\cite{gen_bfkl}
\begin{equation}
\Phi(Q^2,p_V^2)_\text{BFKL} = \exp\left(-
\frac{2 C_A \alpha_s}{\pi w} \; \log \frac{Q}{p_V} \right) 
\left[ 1+ (1-u) \frac{2 C_A \alpha_s}{\pi w} \; \log \frac{Q}{p_V} 
\right]^{u/(1-u)}  \; .
\label{eq:bfkl}
\end{equation}
In this expression $w$ is the Mellin conjugate variable of $x$ which for
the physical jet rates requires convolution with the structure functions and 
transformation back to $x$ space.  In analogy to \eq{eq:gf_evolution}, functional
derivatives of $\Phi_\text{BFKL}$ evaluated at $u=0$ return exclusive
jet rates.\bigskip

As in the DGLAP case of Sec.~\ref{sec:ps} we can compute the scaling
patterns in the limit of small and large emission probability.

\begin{itemize}
\item[(1)] $\dfrac{2 C_A \alpha_s}{\pi w} \log \dfrac{Q}{p_V} \gg 1$\medskip

Following \eq{eq:bfkl} this large logarithm describes the limit of
large emission probabilities. Again, we find that the $\nj$
distribution shows a Poisson scaling
\begin{equation}
\left. \dfrac{1}{n!} \dfrac{\partial^n}{\partial u^n} 
\left[ 1+ (1-u)\dfrac{2 C_A \alpha_s}{\pi w}  \log 
\dfrac{Q}{p_V} \right]^{u/(1-u)} \right|_{u=0}
\; \approx \;
\frac{1}{n!} \log^n \left( 1+
\frac{2 C_A \alpha_s}{\pi w}  \log \frac{Q}{p_V}
\right) \; .
\label{eq:bfkl_po}
\end{equation}
Note that this result is obtained by taking the limit 
after the differentiation.
Although \eq{eq:bfkl_po} is formally true in the limit $1/w \log Q/p_V \to
\infty$ the enhancement of the Poisson term is gradual, 
\ie $\log^n ( 1/w \log Q/p_V )$.  At every multiplicity there 
appear terms of the order $\log^{n-1} ( 1/w \log Q/p_V )$ with 
possibly large coefficients.  Therefore, we expect this Poisson distribution to 
only emerge at very high energies and currently experimentally 
inaccessible $x$ values.

\item[(2)] $\dfrac{2 C_A \alpha_s}{\pi w} \log \dfrac{Q}{p_V} \ll 1$ \medskip

Expanding to leading order in the emission probability the rates are
\begin{equation}
P_{n}
\; \approx \;
\left( \frac{2 C_A \alpha_s}{\pi w}  \log \frac{Q}{p_V}
\right)^n. 
\label{eq:bfkl_stair}
\end{equation}
This is a staircase distribution in the jet ratios.  This result can
also be seen in the corresponding fixed order
computation~\cite{bfkl_rates}.  Apparently, this staircase
distribution is of an entirely different origin from the gluonic
cascade in DGLAP, \eq{eq:gf_staircase}.  In the BFKL evolution there
is no notion of subsequent emissions, all gluons are emitted directly
from the factorized $t$-channel.  A subsequent emission does not
contain the $\log 1/x$ enhancement.  For the physical jet rates, we
transform to $x$ space via a convolution and the rates are of course
different; only the scaling, \ie the shape of the ratio distribution,
does not change.
\end{itemize}
\bigskip

Unlike the patterns we see in the DGLAP approach, both limits pointed
to above are only realized in corners of phase space which are not
particularly relevant to the hard processes at the LHC.  The actual
$\nj$ distributions in BFKL-type regions will not be described in
terms of a pure scaling pattern.  In fact, it was pointed out in
Ref.~\cite{gaussian} that the $\nj$ distribution from simulated BFKL
emissions are described better by a shifted Gaussian than a Poisson
distribution. From a scaling perspective it is difficult to imagine
how this might be induced by first principles QCD.

\section{LHC predictions}
\label{sec:exper}

ATLAS and CMS analyses based on 2011 and 2012 data with an
excellent understanding of the detectors allows for detailed
investigations of high-jet-multiplicity final states. The scaling
patterns derived in this work should hence be experimentally testable
before the 2013 shutdown.\medskip

From the discussion in Sec.~\ref{sec:dy_lhc} we know that $n$-jet
scaling arguments can be transferred from $e^+ e^-$ to hadron
colliders without major modifications. This means we can search for
Poisson scaling and staircase scaling in standard candle processes at
the LHC.  For reasonably large jet multiplicities both of these
patterns are derived from first principles QCD.  Parton density
effects modify the first few bins in the exclusive jet ratios
$R_{(n+1)/n}$.

In the definition of the observables we need to carefully distinguish
between inclusive and exclusive jet rates, \ie counting exactly $n$
jets or at least $n$ jets in the $n$-jet bin radiated off a given hard
processes.  Staircase scaling is unique in the sense that it appears
in exclusive and inclusive observables with the same constant ratio
$R_{(n+1)/n}$~\cite{us_prl}. Poisson scaling at $e^+ e^-$ colliders or
hadron colliders is limited to exclusive jet rates. Its translation
into inclusive observables is tedious~\cite{us_prl}.\medskip

Following Sec.~\ref{sec:dy_lhc} we expect that QCD features together
with the PDF effects produce a convincing staircase pattern for cross
sections with a democratic jet selection and in the absence of major
kinematic cuts~\cite{us_prl,us_photon}. The ratios $R_{(n+1)/n}$
should be constant, \eq{eq:stair}. Towards large jet multiplicities
this constant value $R$ is largely independent of the underlying 
hard process. The first $\nj$ bin is most sensitive to the parton 
densities, as shown in Sec.~\ref{sec:hadron_pdf}. The PDF effect 
will generally lead to a suppression of the otherwise large value 
of $R_{1/0}$, extending the staircase pattern to low multiplicities 
where the QCD derivation fails. However, the PDF effect is clearly 
not independent of the hard core process which can involve incoming 
gluons as well as incoming quarks.

In contrast, for a jet selection with a large ratio between the
core-process scale and the jet acceptance cut ($p_{T,j}\ge p_V$) we
expect several bins of the $\nj$ distribution to follow a Poisson
distribution. Only at large jet multiplicities the ratios will
overshoot the Poisson pattern $\bar{n}/(n+1)$, turning into a
staircase tail. This is the same behavior we derive for $e^+ e^-
\to$~jets production in Sec.~\ref{sec:it_poisson}.\medskip

Our considerations about the $\nj$ distributions associated with a
general hard process needs to be validated for a variety of channels,
potentially in dedicated analyses. As a starting point, we propose 
two simple extensions of already existing analyses.  They serve as 
stringent tests in particular of the Poisson hypothesis for the 
low-multiplicity bins and the transition to a staircase-like 
behavior for higher $\nj$ bins. We start with the Drell-Yan process, 
but asking for a leading jet with significant transverse momentum 
$p^\text{lead}_T \ge 5 p_V$ while all additional jets are selected 
democratically with $p_T\ge p_V$. Our second example is the $\nj$ 
distribution inside a rapidity interval defined by a dijet core system.

\subsection{Z+jets}  

For the $W/Z+$jets channels recent LHC measurements of the $\nj$
distribution support a staircase
hypothesis~\cite{ex_lhc,atlas_zjets}. The key ingredient is the
democratic selection of jets, \eg $p_{T,j} > 30$~GeV for all jets in
$Z \to \mu^+ \mu^-$ and $Z \to e^+ e^-$ events~\cite{atlas_zjets}. We
propose a minimal modification of this analysis by increasing the
$p_T$ cut on the leading jet to $p^\text{lead}_T > 150$~GeV. The hard
core process is now defined as the $Z$ boson combined with the leading
jet, the first of $\nj$ radiated QCD jets is the second jet in the
event.  According to Sec.~\ref{sec:hadron_pdf} such a selection makes
the PDF suppression effect rather marginal and we expect to see a
Poisson distribution in the low-multiplicity bins.\medskip

\begin{figure}[t]
\centering
\includegraphics[width=0.495\textwidth]{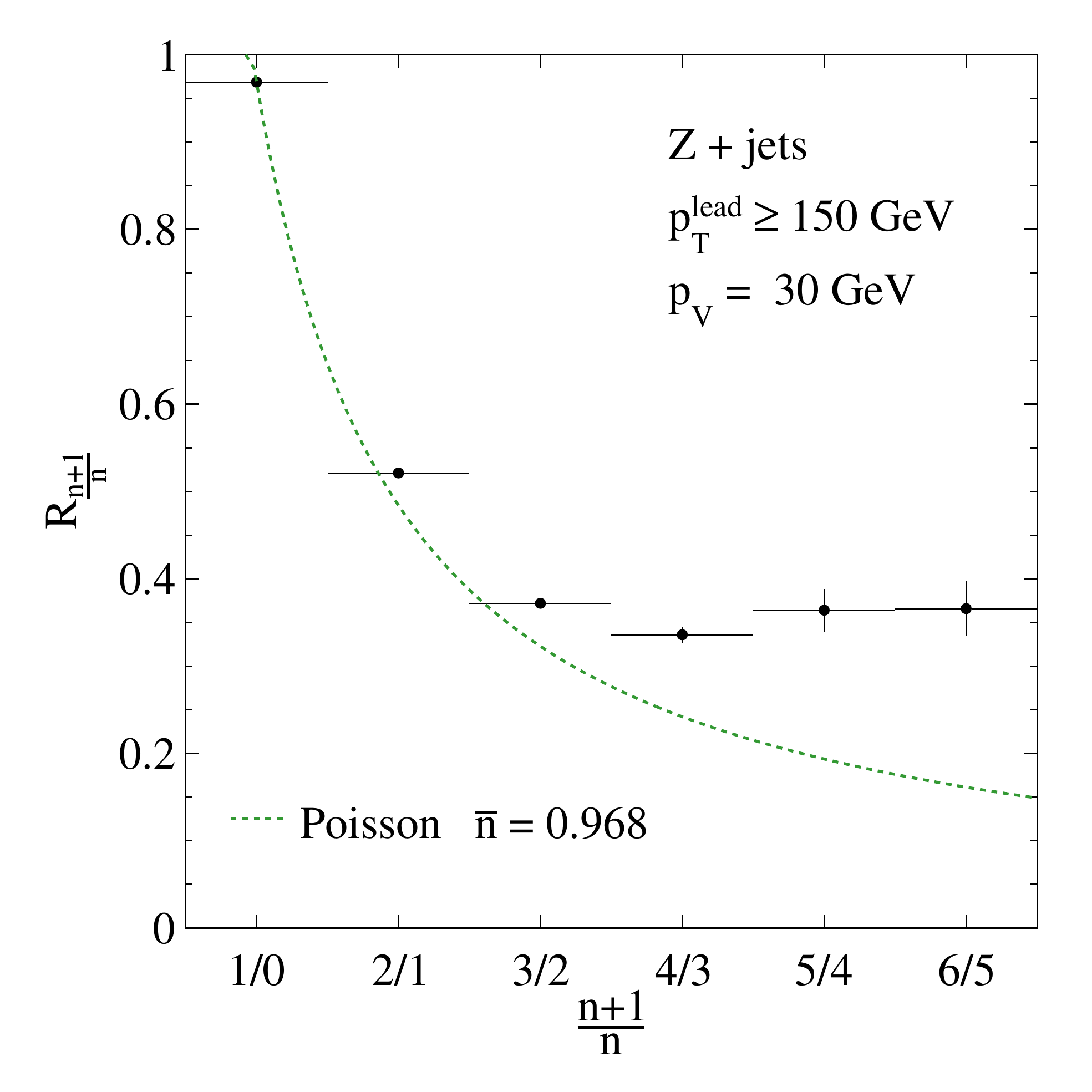}
\hfill
\includegraphics[width=0.495\textwidth]{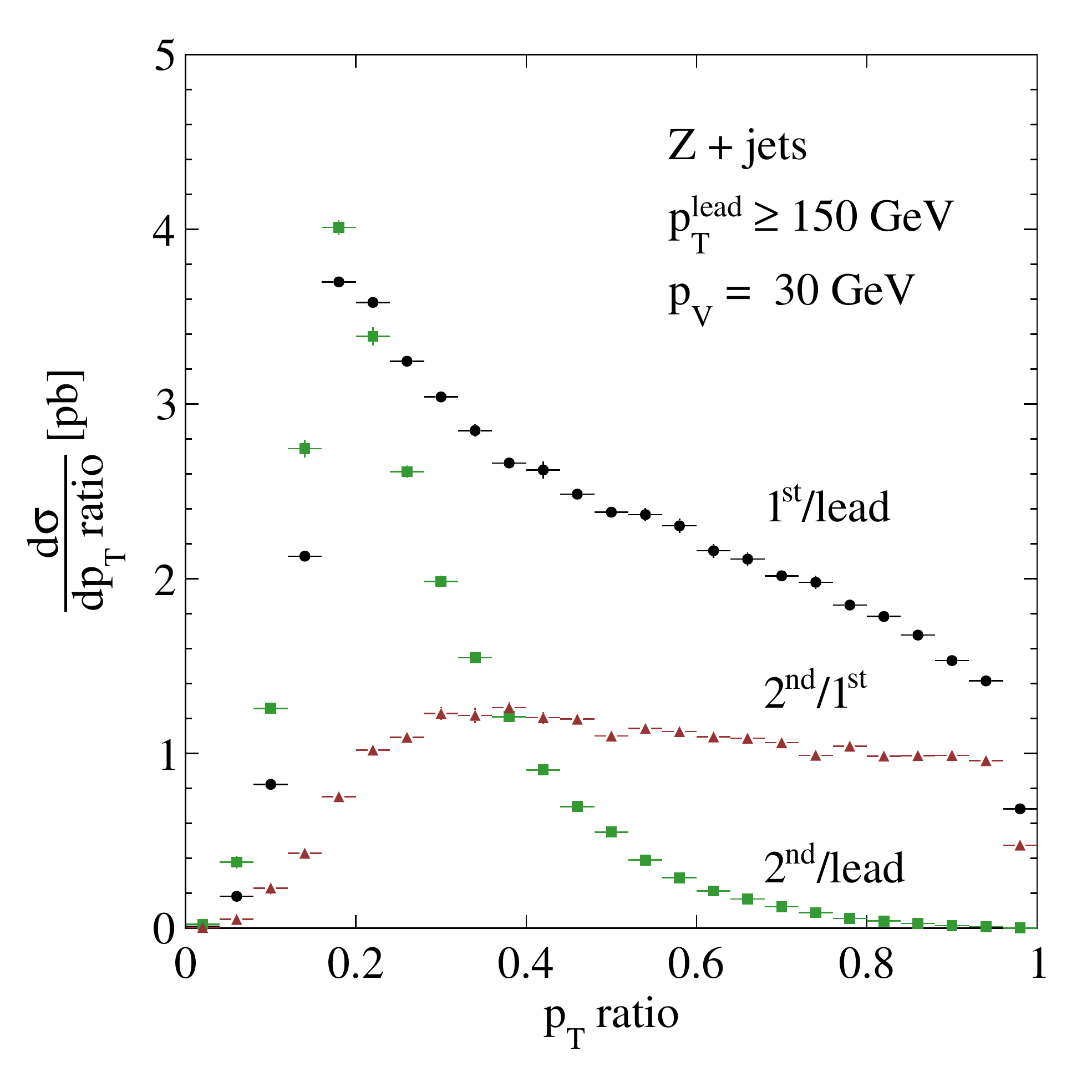}
\caption{Left panel: exclusive jet ratios for $Z\; +\;$jets production
  at $\sqrt{S}=7$~TeV.  We require a leading jet with $p^\text{lead}_T
  \ge 150$~GeV. All other jets have $p_T\ge p_V=30$~GeV. The line
  shows a Poisson shape with $\bar{n}$ extracted from the first
  bin. Right panel: cross sections as a function of the event-wise
  ratios of the jet transverse momenta in inclusive $Z + 3$ jet
  production.  While the leading jet has to pass $p^\text{lead}_T\ge
  150$~GeV additional jets are selected uniformly with just $p_T\ge
  p_V=30$~GeV.}
\label{zjets_nj}
\end{figure}

To simulate this process we use Sherpa
v1.4.0~\cite{sherpa,sh_shower}. We employ Sherpa's tree-level merging
algorithm based on truncated showers~\cite{sh_merging}, including
real-emission matrix elements for up to five final state partons, the 
merging scale we set to $25$~GeV.
In \fig{zjets_nj} we present the exclusive $\nj$ cross section ratios,
keeping in mind that the counting of the radiated jets does not
include the hard leading jet which together with the $Z$ boson
constitutes the core process.  As expected, we observe a clear Poisson
scaling for $\nj \le 3$. This can be seen when comparing the
individual bins to the Poisson shape $R_{(n+1)/n} = \bar{n}/(n+1)$
where $\bar{n}$ is fixed by the first bin $R_{1/0}$.

However, already the three-jet rate comes out higher than the Poisson
extrapolation. At this point the non-abelian nature of QCD radiation
takes over, giving us a staircase tail with $R\approx 0.36$. This
behavior is strongly reminiscent of \fig{fig:ee_staircase} for $e^+
e^- \to$~jets. An updated analysis of the 2011 data sample and
certainly the 2012 data set should allow for a test of up to five
additional jets, probing QCD predictions for the complete $\nj$
distribution.\bigskip

In addition to its defining power of very general features of the
$\nj$ distribution the jet selection cuts are reflected in the
transverse momentum distributions of the jets.  In the right panel 
of \fig{zjets_nj} we depict the cross sections as a function of the 
ratios of the $p_T$-ordered jet transverse momenta defined event-by-event. 
For this distribution we for once deviate from our usual exclusive 
jet counting and consider events with at least three jets in the final
state. We consider the $p_T$ ratios first-over-leading (black), 
second-over-leading (green) and second-over-first jet (red). With the 
leading jet $p_T \ge 150~$GeV and the additional jets selected 
uniformly with $p_T \ge 30~$GeV we expect the first and second jet 
to peak around the selection cut. This is confirmed by the simulated 
results that exhibit strong peaks for the corresponding ratios 
around $p^{\rm 1st}_{T}/p^{\rm lead}_{T}\approx 0.2$ and
$p^{\rm 2nd}_{T}/p^{\rm lead}_{T}\approx 0.2$.  However, it is interesting 
to note that the ratio $p^{\rm 2nd}_{T}/p^{\rm 1st}_{T}$ does not 
peak around $1$. Rather QCD favors the first radiated jet to be 
significantly harder than the second. The obtained distribution in 
fact turns out to be more or less flat between $0.25$ and $1$. It 
is certainly interesting to study these observables in addition to 
the $\nj$ distribution, as they contain complementary information 
on the underlying QCD dynamics.

\subsection{QCD gap jets} 
\label{sec:exper_gaps}

\begin{figure}[t]
\centering
\includegraphics[width=0.45\textwidth]{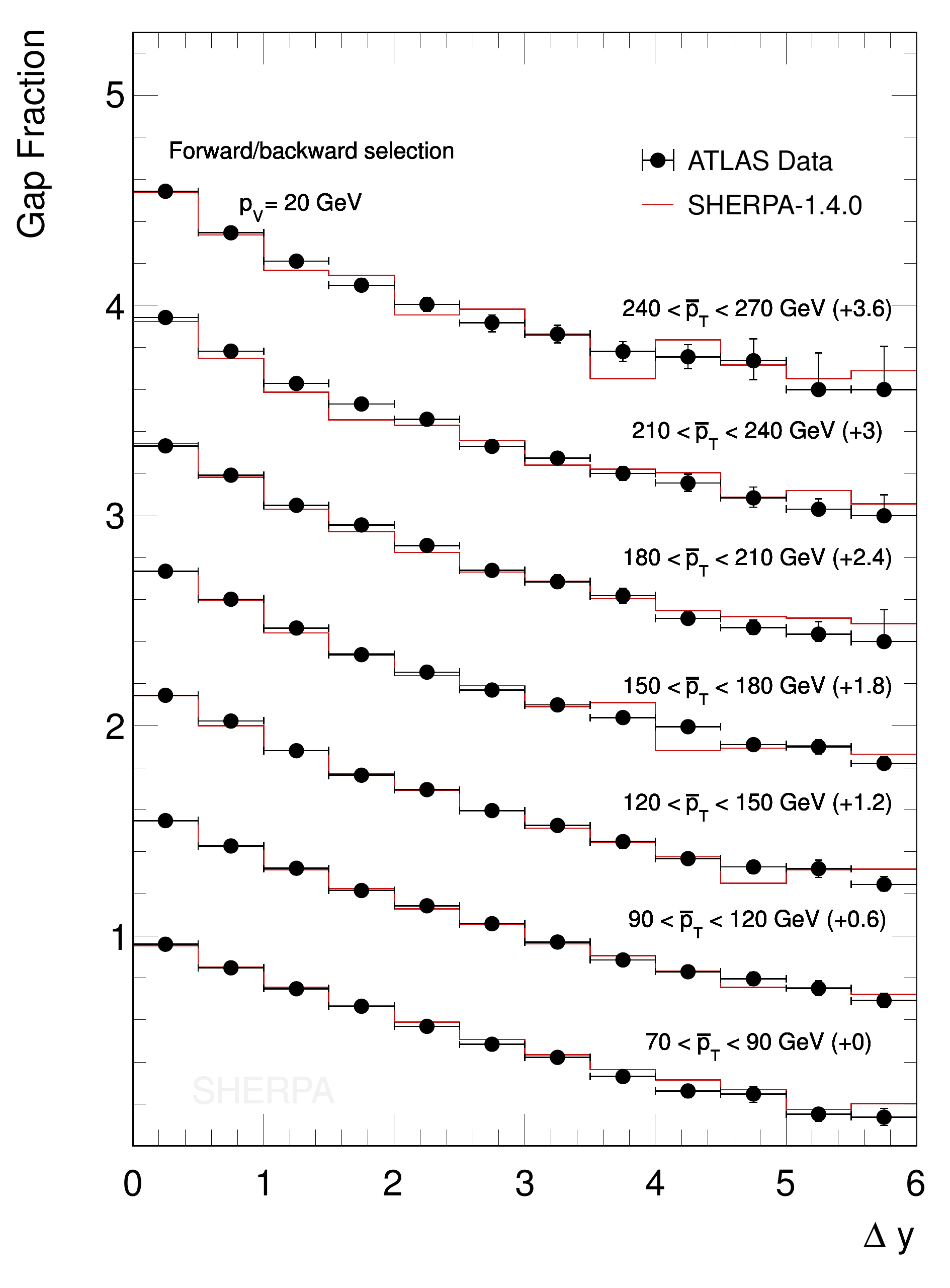}
\hspace{.5cm}
\includegraphics[width=0.45\textwidth]{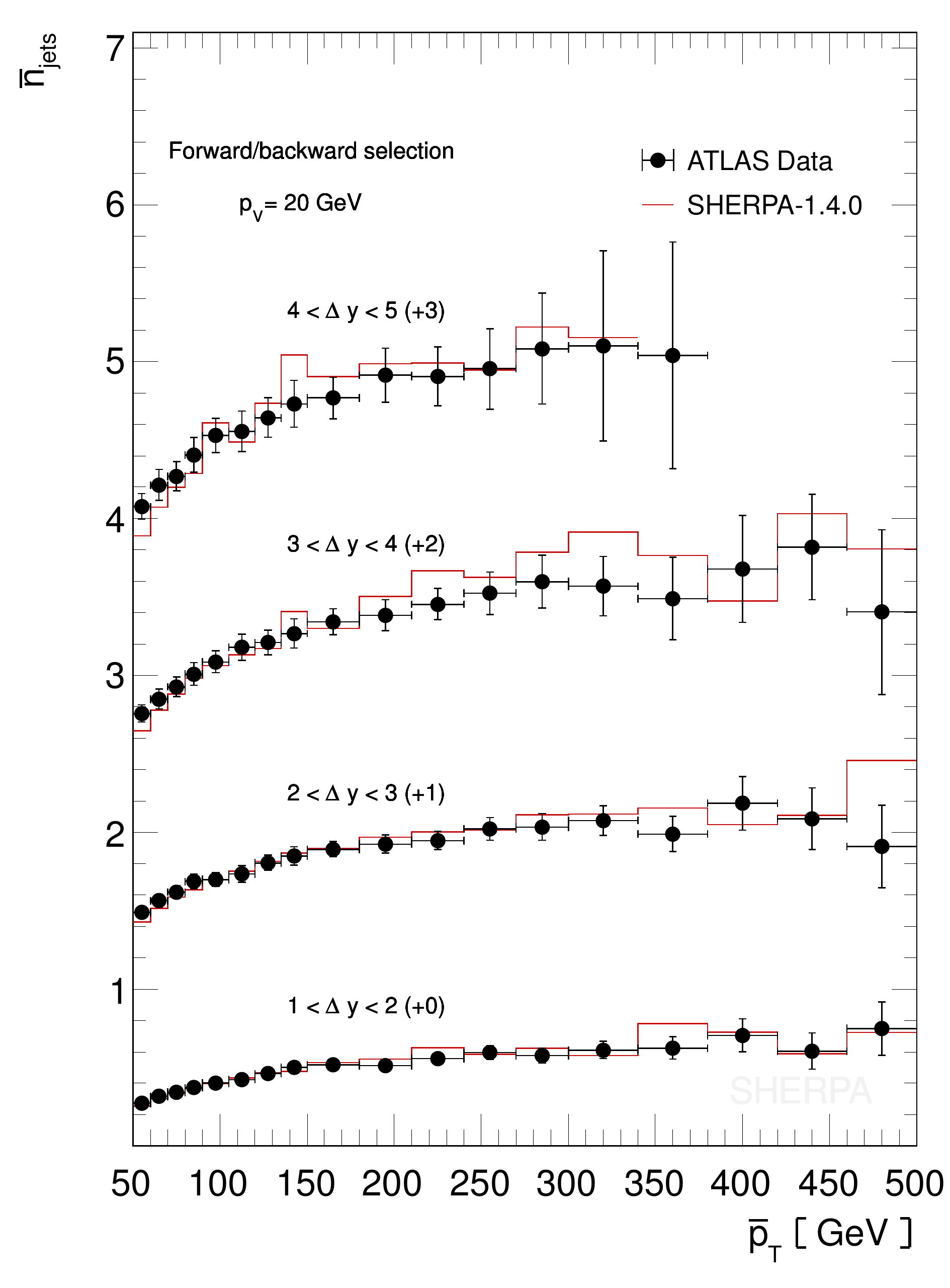}
\caption{Left: Predictions vs ATLAS data~\cite{gapjets} for the di-jet
  gap fraction ($p_V=20$~GeV) as a function of $\Delta y$ and in
  slices of $\bar{p}_T$. Right: average number of gap jets as a
  function of $\bar{p}_T$ in slices of $\Delta y$. The tagging jets
  are defined as most forward and most backward. All jets are
  reconstructed using the anti-$k_t$ algorithm with $R=0.6$.}
\label{fig:gap_ob}
\end{figure}

An interesting set of observables from the perspective of multi-jet
final states are gap fractions or gap jets. In that case we require a
specific kinematic structure of hard and widely separated jets and
count the QCD jets in between. The core process is the production of
two widely separated hard jets.

A recent ATLAS study~\cite{gapjets} identifies two forward jets,
so-called tagging jets, either as the highest $p_T$ (selection A) or
the most forward and backward in rapidity (selection B).  The core
di-jet system is defined in terms of $\bar{p}_T = (p_{T,1} +
p_{T,2})/2$ and $\Delta y = |y_1 - y_2|$.  The gap fraction $P_0 =
\sigma_0/\sigma_\text{tot}$ is given by all events with no additional
jet in between the two tagging jets and $p_T > p_V = 20$~GeV.  ATLAS
measures it as a function of $\Delta y$.  In addition, they provide us
with the average multiplicity of vetoed jets or gap jets.  Following
Sec.~\ref{sec:it_poisson} a hard cut on $\bar{p}_T$ will typically
enforce a larger logarithm than simply $\log(\bar{p}_T/p_V)$. This is
because the two tagging jets tend to be asymmetric in $p_T$.  As a
consequence we expect a Poisson scaling for the gap jets.

Perturbatively, the gap fractions are dominated by the single emission
probability while the average number of gap jets probes multiple
emissions.  We again propose a simple extension of the existing
analysis. In addition to the average number of gap jets the
corresponding exclusive $\nj$ distribution should be
studied. Reproducing the average number of gap jets cannot validate a
full $\nj$ distribution, where the latter directly tests QCD scaling
patterns.\bigskip

Again, we use Sherpa with tree-level merging, including hadronization
and underlying event. We account for up to four hard final state
partons in the veto region, the merging scale set to $Q_\text{cut}=20$~GeV. 
This way, the emission of wide-angle soft gluons should be correctly
modeled including the full color structure. The DGLAP based parton 
shower resums logarithms of large $p_T$ ratios only, for sufficiently 
large rapidity gaps one may expect the agreement to suffer~\cite{hej}.  
In \fig{fig:gap_ob} we compare Sherpa predictions with ATLAS
data~\cite{gapjets} for both the gap fraction and the average number
of jets in selection B.  For both observables the agreement with data
is excellent. Even for large $\Delta y > 4 $ the simulation works well 
within the statistical limitations, so we do not see the merging of 
tree-level matrix elements with the parton shower
breaking down. A detailed analysis of the same data using the MC@NLO
implementation in Sherpa and assessing both perturbative and
non-perturbative theoretical uncertainties is presented in
Ref.~\cite{sh_jets}.\medskip

\begin{figure}[t]
\centering
\includegraphics[width=0.45\textwidth]{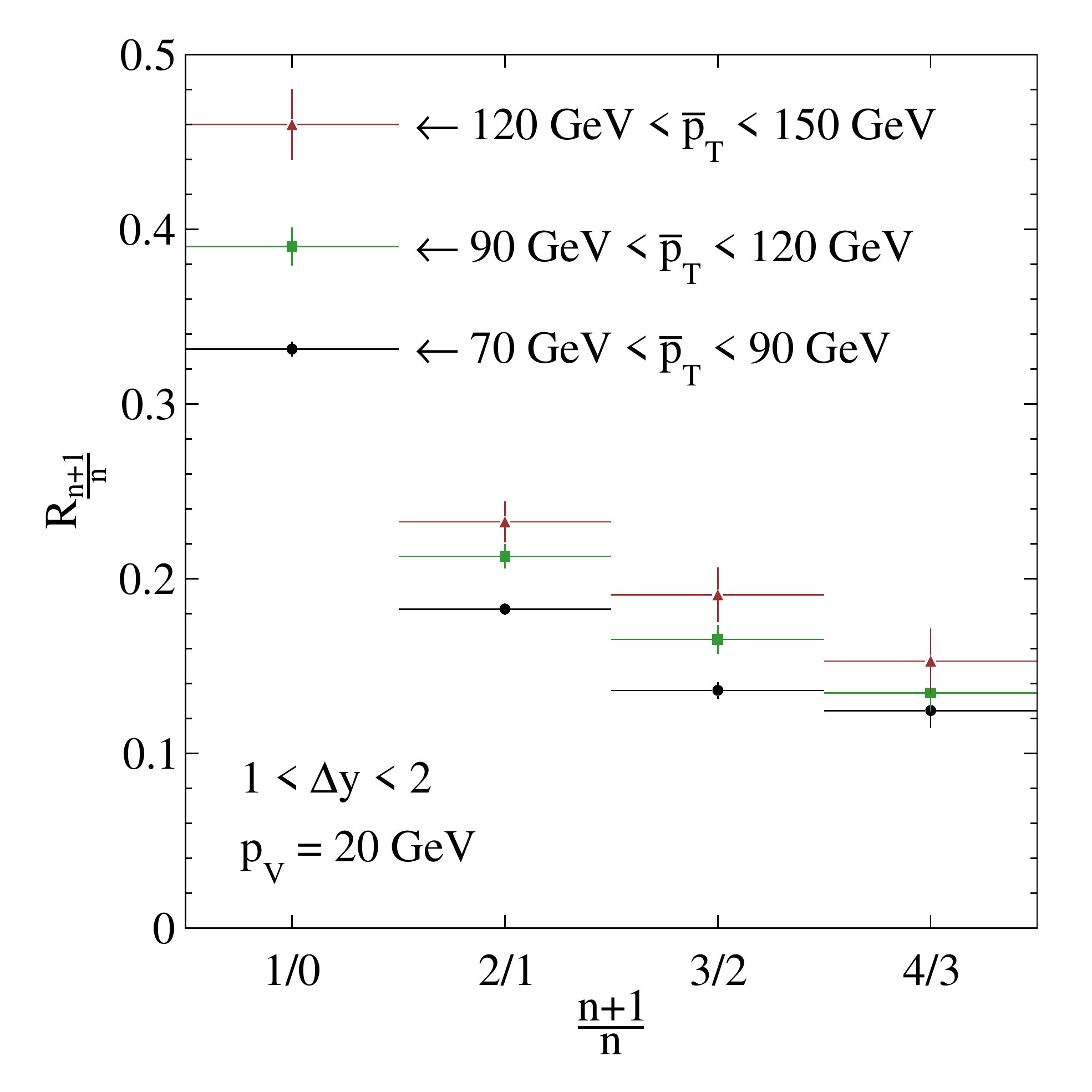}
\hspace{.5cm}
\includegraphics[width=0.45\textwidth]{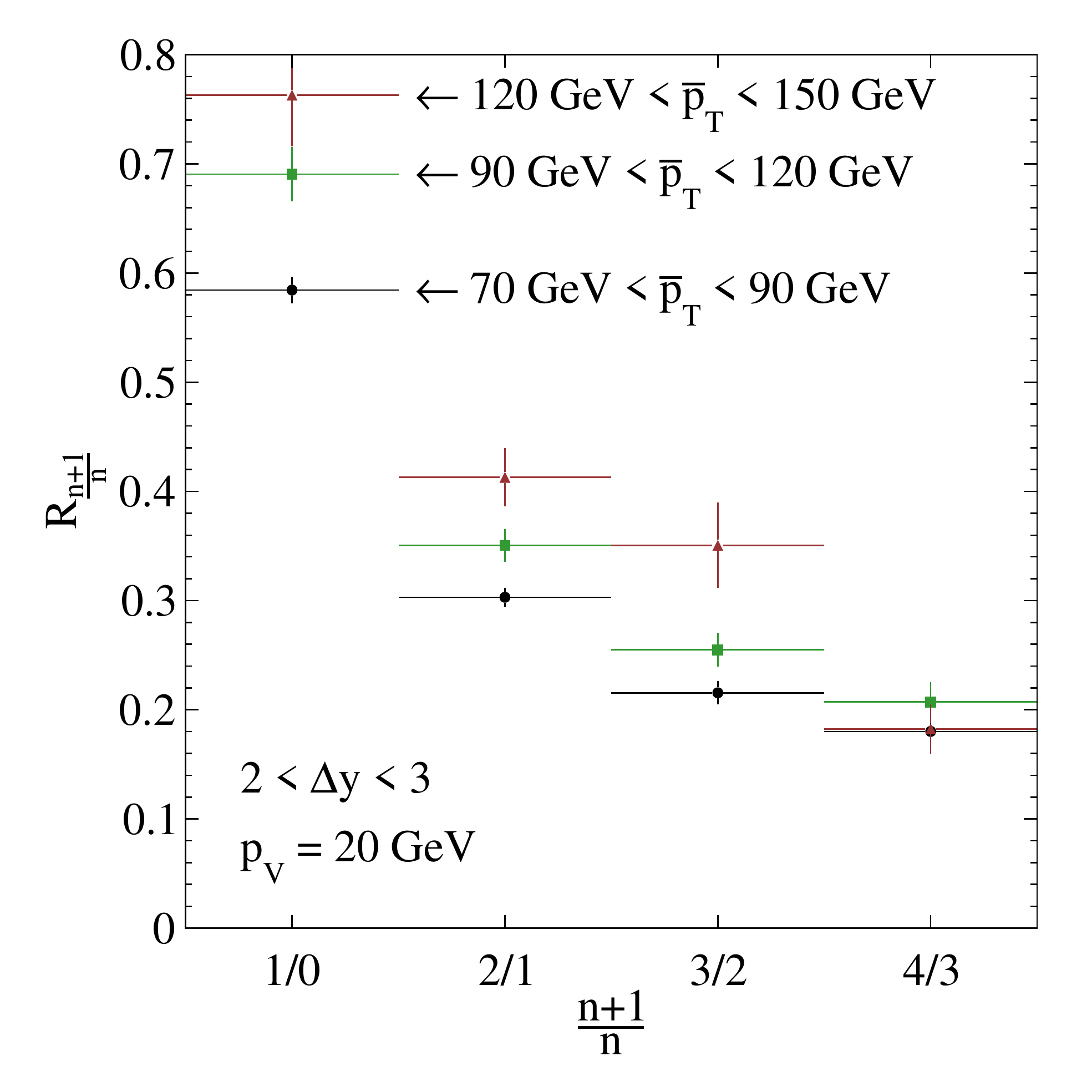}
\caption{Jet ratios for exclusive gap jets in slices of $\bar{p}_T$
  for the tagging jets, defined as the most forward and backward
  in the event. The rapidity separation of the gap-defining
  jets is $1< \Delta y < 2$ (left) and $2 < \Delta y <3$ (right). }
\label{fig:njet_gap}
\end{figure}

With the confidence gained by correctly modeling the gap fraction and
the average number of gap jets, we proceed to predict the $\nj$
distribution for gap jets.  In the forward-backward selection this
includes all resolved jets with $p_T > p_V$.  In \fig{fig:njet_gap} we
present the gap jet multiplicity ratios for two different rapidity
separations, $1< \Delta y<2$ (left) and $2< \Delta y<3$ (right), 
and in slices of $\bar{p}_T$. Enforcing a large ratio between the 
$p_T$ of the tagging jets versus the jet-counting scale $p_V = 20$~GeV 
induces a Poisson distribution for all the considered $\bar{p}_T$ 
selections.  Comparing the two panels of \fig{fig:njet_gap} we see 
that the shape, though not the normalization, of the $\nj$ 
distributions is largely independent of $\Delta y$. For large 
$\bar{p}_T$ the highest jet multiplicities suffer from 
non-negligible statistical fluctuations, in particular for the 
$4/3-$jet bin.

For kinematical selections which induce a Poisson scaling we can
approximately derive the average number of gap jets from
$R_{1/0}$. This number determines $\bar{n}$ and allows for a 
straightforward calculation of the veto survival probability or 
gap fraction, $P_0=\exp(-\bar{n})$.  Comparing this prediction to 
the explicit $\bar{n}$ measurement of in \fig{fig:gap_ob} we find 
both values consistent. Such a comparison will provide insight into 
jet-veto survival probabilities, not only for the analysis presented 
here but also in the context of Higgs analyses~\cite{us_prl,cjv}.

\section{Conclusions}
\label{conclusions}

While it has been known for a long time that the exclusive $\nj$
distribution at hadron colliders follows simple scaling
features~\cite{scaling_orig}, a proper understanding starting from
first principles QCD has been missing. The two underlying scaling
patterns for jet emission is a Poisson shape or staircase scaling. 

\begin{enumerate}
\item Poisson scaling is well known from multi-jet final states in
  $e^+ e^-$ production. It can be easily derived in the abelian limit
  of QCD, \ie gluon radiation off hard quark legs, in complete analogy
  to soft photon emission in QED. Using the parton shower picture as
  well as generating functionals we have found that a Poisson distribution
  is tied to a large hierarchy of scales and a logarithmically
  enhanced radiation matrix element. Such different scales can be
  induced by kinematic cuts, for example in $Z$+jets~\cite{us_prl} or
  $\gamma$+jets~\cite{us_photon} production.  Many effects, like
  non-abelian gluon splittings, phase space limitations, or
  sub-leading logarithms modify the pure Poisson shape of the
  exclusive $\nj$ distribution.
\item Staircase scaling~\cite{scaling_orig} is defined as constant
  exclusive jet ratios $R_{(n+1)/n} = \sigma_{n+1}/\sigma_n = R$. It
  has been observed at hadron colliders since UA1 and generally
  appears for a democratic jet selection, avoiding large scale
  separations. This includes Higgs production in gluon fusion and in
  weak boson fusion~\cite{us_prl}. For large jet multiplicities we
  have found that staircase scaling can be predicted as a non-abelian
  QCD effect. At $e^+ e^-$ colliders as well as at hadron colliders it
  describes the exclusive $\nj$ distribution for large jet
  multiplicities. At lower jet multiplicities parton density effects
  violate our picture of jet radiation off an unchanged hard
  process. Numerically, this correction restores the staircase
  pattern, often down to the first emitted jet.
\end{enumerate}

After this derivation and study of the two generic patterns which are
based only on QCD and patterns of the parton densities this prediction
should be tested experimentally. We present two minor modifications of
current LHC analyses which would be well suited to probe the universal
nature of our findings. First, the classical $Z$+jets channel should
be used to study the transition from Poisson to staircase scaling
using a simple kinematic cut on the leading jet. A similar proposal
exists for $\gamma$+jets production~\cite{us_photon}. In addition, QCD
gap jets already probe these scaling features through the measured
average number of gap jets. Extending the main observable from the 
mean number of gap jets to the $\nj$ distributions would allow us to
confirm what our first estimate suggests --- that gap jets can be
described precisely using the transition from staircase to Poisson
scaling.\medskip

While from a perturbative QCD point of view the exclusive number of
jets $\nj$ is not a particularly attractive observable, it clearly has
huge benefits in LHC analyses. In the future, jet vetoes in Higgs
searches~\cite{us_prl} are going to become more and more relevant,
separating the different underlying production processes. In this
paper we have shown that it is possible to predict the main features
of the $\nj$ distribution from QCD. Testing the universality of these
features would give us a new and improved tool for a huge number of
LHC analyses to come.

\bigskip
\acknowledgments We would like to thank Bryan Webber, Einan Gardi,
Jennifer Smilie, Patrick Meade and Peter Zerwas for enlightening
discussions as well as the G\"ottingen ATLAS group for experimental
input. Moreover, we are grateful to Michelangelo Mangano for setting
this problem.  EG and SS acknowledge financial support by the
Helmholtz Alliance 'Physics at the Terascale' and BMBF under contract
05H12MG5. PS acknowledges support by the IMPRS for Precision Tests of
Fundamental Symmetries.

\newpage

\appendix 


\section{Iterated Poisson process}
\label{app:toy}

The process generated by summing an arbitrary number 
of independent Poisson processes $P = P(\bar{n}_1)  +  P(\bar{n}_2) 
+\, \cdots \, P(\bar{n}_n)$ is itself a Poisson process 
with the expectation value $\bar{n} = \bar{n}_1 + \bar{n}_1 + \cdots \bar{n}_n$.    
In order to deviate from a Poisson shape, we allow for the occurrence of
subsequent splittings. A QCD example would be gluon radiation off a
quark line with a subsequent splitting of the gluon into two gluons or
two quarks.

\begin{figure}[t]
\centering
\includegraphics[height=0.49\textwidth]{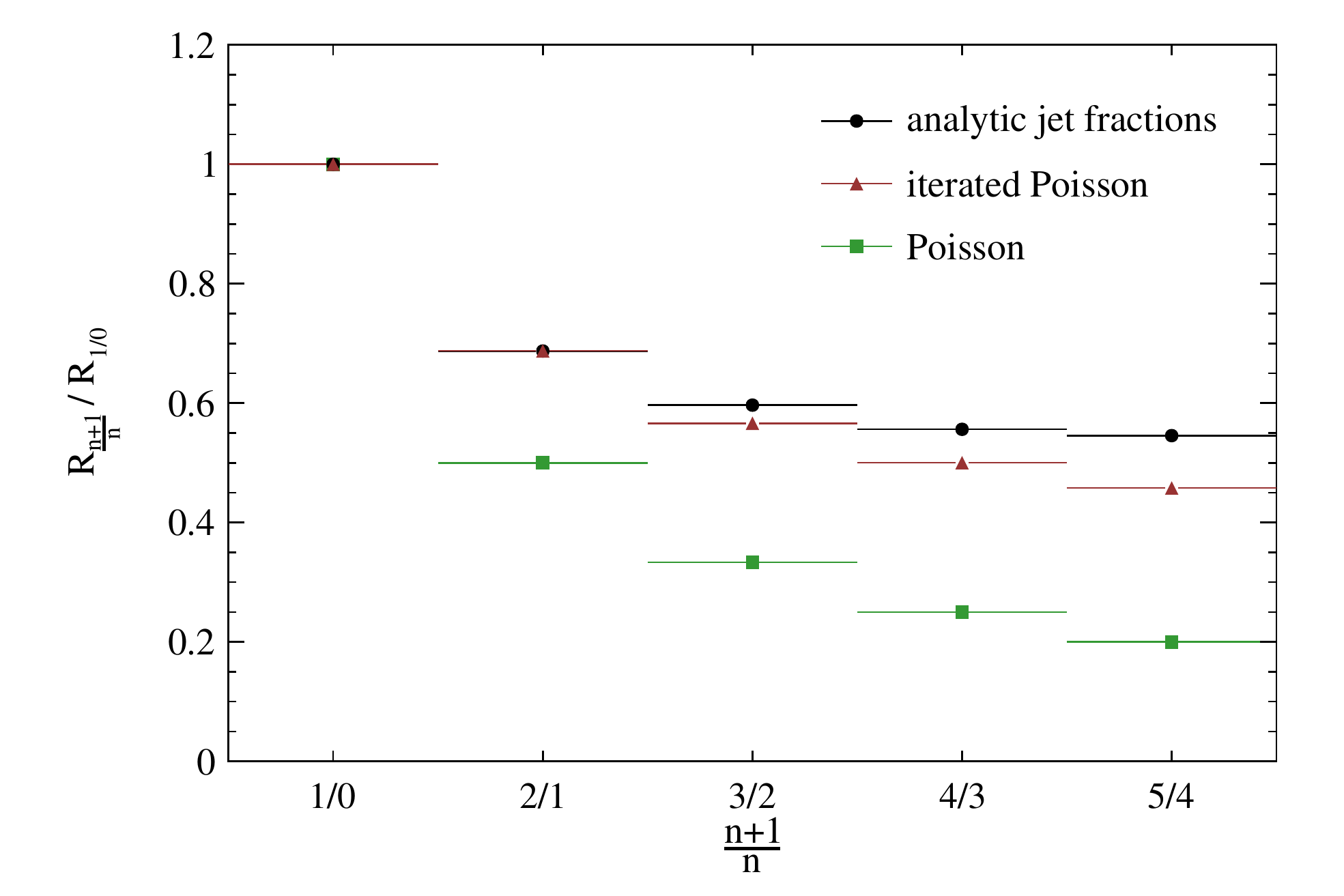}
\caption{Ratios of $e^+ e^- \to q\bar{q} \,+ \,\text{n} g$ jet
  fractions (black) as computed in Appendix~\ref{app:fractions}
  compared to the iterated Poisson model with $\bar{n}' = 12C_F/C_A \,
  \bar{n}$ (red), and Poisson scaling (green).}
\label{fig:cuba_libre}
\end{figure}

Suppose we start with a single mother Poisson process described by the 
parameter $\bar{n}$.  Every emission by the mother process generates 
a daughter process which itself evolves through the same scale as an 
independent Poisson process with expectation $\bar{n}'$.  In 
general we will have $\bar{n} \ne \bar{n}'$.  
For the mother (daughter) process with Poisson parameter $\bar{n}$
$(\bar{n}')$, the two parameter distribution can be approximated by
\begin{alignat}{5}
P(n;\bar{n},\bar{n}') 
& = \; e^{-\bar{n}-n \bar{n}'}\displaystyle{\sum_{i=0}^{n}}	
\left( \frac{(n-1)!}{i! (n-i-1)! (n-i)! } \right) \bar{n}'^{i} \bar{n}^{n-i} 
\notag \\
&  = \; \frac{e^{-\bar{n} - n \bar{n}' } \, \bar{n}^n \; \bar{n}'^{n-1}}{n!}
\left|\,\text{HG}\left[1-n\, ,2\, ,{-\bar{n}}/{\bar{n}' }\right] \right| \; ,
\label{eq:aug_pos}
\end{alignat}
where HG is the confluent hypergeometric function with integral
representation
\begin{equation}
\text{HG}[a,b,z] = \frac{1}{\Gamma(a)} 
\int_0^{\infty} e^{-zt} t^{a-1}(1+t)^{b-a-1}dt \, .
\end{equation}
Let us note some observations regarding the physics content of this model:
\begin{enumerate}
\item The exponential Poisson model is reproduced in the limit
  $\bar{n}' \to 0$, where the additional splitting probability
  vanishes.

\item The zero emission probability is unchanged, while the one
  emission rate includes the additional non-splitting probability of
  the daughter process. It contains an additional suppression
  $e^{-\bar{n}'}$ which universally reduces the first $n$-jet ratio
  for the iterated Poisson process.

\item In the limit of a large additional splitting probability
  $\bar{n}' \gg \bar{n}$ the theory displays staircase scaling.  The
  highest order term in \eq{eq:aug_pos} carries a compensating factor
  of $n!$ in the numerator.

\item The convergence of \eq{eq:aug_pos} is guaranteed for all
  $\bar{n}$ and $\bar{n}'$.  However, inclusive unitarity of the
  cross-section is not possible; the sum over $n$ in \eq{eq:aug_pos}
  is smaller than one for all $\bar{n}' \ne 0$.  We expect then that
  our toy model undershoots the full calculation, as confirmed by 
  the $e^+ e^-$ jet fractions (see \fig{fig:cuba_libre}).
 
\item We emphasize that the purely probabilistic model given by 
  \eq{eq:aug_pos} does not capture all effects of leading-log QCD.  
  In order to promote our toy model to full double leading 
  logarithmically accurate QCD as far as the jet rates are concerned, 
  we need to make two modifications. First of all we need to include 
  the full Poisson history by hand (\ie no recursive formula is 
  achievable), and second, we require $\bar{n}'$ to change as a 
  function of the ``generation" of subsequent emissions.  More precisely 
  the fully non-abelian $\sim C_F C_A^2$ three-gluon rate is correctly 
  described if $\bar{n}'' =  (6/5) \bar{n'} $.  This value for 
  $\bar{n}''$ then correctly describes the $C_F^2 C_A^2 $ coefficient 
  to the four-gluon rate.  In this manner we can essentially boot-strap 
  the leading coefficients.   
  
\end{enumerate}

\begin{figure}[t]
\centering
\includegraphics[height=0.49\textwidth]{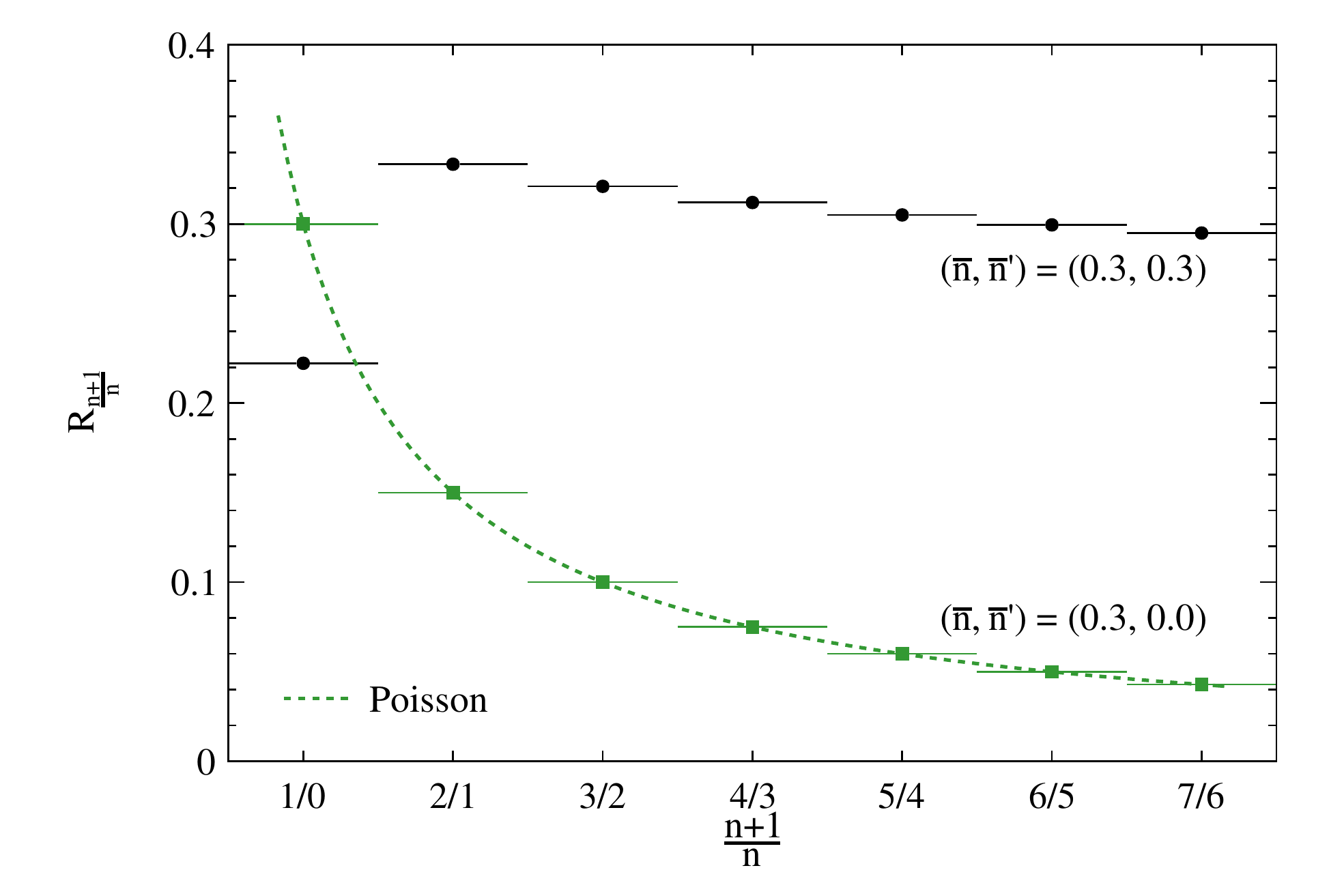}
\caption{Ratios from an iterated (normal) Poisson process for 
$\bar{n} = 0.3 $ and  $\bar{n}'=3$ ($\bar{n}'=0$) given in \eq{eq:aug_pos}.  
In the iterated case we see a general flattening of the distribution. } 
\label{fig:aug_p_pl}
\end{figure}

The important consequence of \eq{eq:aug_pos} for realistic values of
$\bar{n}'>0$ is that the jet ratios $R_{(n+1)/n}$ are flatter than the
naive Poisson expectation.  In \fig{fig:aug_p_pl}, we display the jet
ratios generated by \eq{eq:aug_pos} for $\bar{n} = \bar{n}'$.  By making 
analytic estimates for the subsequent splitting parameter we can 
compare our toy model with jet rates predictions from QCD.

We assume the relation between the two expectation values $\bar{n}$ 
and $\bar{n}'$ to scale with the
relative splitting probability for the simplest non-abelian
emission. The $2$-jet fraction at $\ope(a^2)$ in double logarithmic approximation then gives
$\bar{n}/ \bar{n}' = 12 C_F/C_A$.  For $\bar{n} < 1$, the specific
value of the two Poisson parameters do not affect the shape as long as
their ratio is held fixed.  Therefore, we compare the shapes of the
iterated Poisson model with the jet fractions by normalizing the
first ratio.  This also allows a simple comparison to a purely Poisson
process with ratios $1/(n+1)$.  The results are shown in
\fig{fig:cuba_libre}, where we find that the $P_2/P_1$ ratios are in
exact agreement.  Using \eq{eq:aug_pos} the ratio is simply $\bar{n}/2 +
\bar{n}'$, which our choice of $\bar{n}'$ reproduces 
precisely.  The higher bins slightly undershoot the jet fractions in the
iterated Poisson model. The reason stated previously is that the $\bar{n}'$ 
should change as a function of the ``generations" and the certain 
contribution need to be added by hand.  The first evidence of this is contained
at $\ope(a^3)$ in $P_3$.  The jet fraction coefficients for the
$\bar{n}\bar{n}'$ and $\bar{n}'^2$ terms are $1/12$ and $1/90$
respectively, while the iterated Poisson model gives 
$1/12$ and $1/144$.  For higher terms still the discrepancy
persists.  The conclusion from this plot is that the purely final state 
gluon cascade generates a staircase scaling pattern at higher 
multiplicities but seemingly fails to explain the suppression on 
the lower multiplicity bins.  For this we need to account for PDFs. 

\section{Jet fractions}
\label{app:fractions}

We briefly quote $e^+ e^- \to q\bar{q} \, + n g$ jet fractions 
at $\ope(\alpha_s^5 L^{10})$ in double logarithmic approximation and for the Durham algorithm.  These are used to compare
the relative size of subsequent terms in a purely final state cascade picture where 
the single emission probability is still smaller than unity.  Results can be obtained from 
Refs.~\cite{av_num_jets,new_clust,leder_jf,webber_jet_rates}.  The abelian contributions 
are shown here to all orders such that in the limit $C_A \to 0 $ we recover Poisson scaling.  
Defining $a=\alpha_s/\pi$, $L = \log y_\text{cut}$ and $P_i = \sigma_i/\sigma_\text{tot}$ 
we have: 

{\small
\begin{alignat*}{5}
P_0 \; & = \; 
            \exp{\left[ -\frac{a C_F L^2}{2} \right]}  \notag \,,\\
P_1 \; & = \;   
        \left(\frac{a C_F L^2}{2}\right)  \exp{\left[ -\frac{a C_F L^2}{2} \right]}  
        \;- 
        a^2 \left ( \frac{C_F C_A}{48} \right)L^4 +
       a^3 \left( \frac{C_F^2 C_A}{96} + \frac{C_F C_A^2}{960}\right) L^6 
       \notag \\ & -      
      a^4 \left( \frac{C_F^3 C_{A}}{384}  
       + \frac{C_F^2 C_{A}^2}{1920} + \frac{C_F C_{A}^3}{21504}\right) L^8 
       \notag \\ & +      
      a^5 \left( \frac{C_F^4 C_{A}}{2304}  
 + \frac{C_F^3 C_{A}^2}{7680} + \frac{C_F^2 C_{A}^3}{43008}
 + \frac{C_F C_{A}^4}{552960}\right) L^{10} \,, \\
%
P_2  \; & = \;  
            \dfrac{1}{2!}\left(\dfrac{a C_F L^2}{2}\right)^2 \exp{\left[ -\frac{a C_F L^2}{2} \right]}
+ a^2 \left( \frac{C_F C_A}{48} \right)L^4 
- 
  a^3 \left( \frac{C_F^2 C_A}{48} + \frac{7 C_F C_A^2}{2880}\right) L^6 
\notag \\  &+  
  a^4 \left( \frac{C_F^3 C_{A}}{128} + 
\frac{C_F^2 C_{A}^2}{512} + \frac{C_F C_{A}^3}{5120} \right) L^8 
\notag \\ & -      
      a^5 \left( \frac{C_F^4 C_{A}}{576}  
 + \frac{ 31 C_F^3 C_{A}^2}{46080} + \frac{23 C_F^2 C_{A}^3}{161280}
 + \frac{C_F C_{A}^4}{806400}\right) L^{10} \,,
\end{alignat*}

\begin{alignat*}{5}
P_3 \; & = \;  
      \dfrac{1}{3!}\left(\dfrac{a C_F L^2}{2}\right)^3 \exp{\left[ -\frac{a C_F L^2}{2} \right]}
+ a^3 \left( \frac{C_F^2 C_A}{96} + \frac{C_F C_A^2}{720}\right) L^6
\notag \\ &-  a^4 \left( \frac{C_F^3 C_{A}}{128}  + \frac{3 C_F^2 C_{A}^2}{1280}
+ \frac{41C_F C_{A}^3}{161280}\right) L^8 
\notag \\ & +      
      a^5 \left( \frac{C_F^4 C_{A}}{384}  
 + \frac{ 19 C_F^3 C_{A}^2}{15360} + \frac{115 C_F^2 C_{A}^3}{387072}
 + \frac{13 C_F C_{A}^4}{460800}\right) L^{10} \notag \,,\\
P_4 & = \; 
            \dfrac{1}{4!}\left(\dfrac{a C_F L^2}{2}\right)^4 \exp{\left[ -\frac{a C_F L^2}{2} \right]}
         +
a^4 \left( \frac{C_F^3 C_{A}}{384}  + \frac{7 C_F^2 C_{A}^2}{7680} 
+ \frac{17C_F C_{A}^3}{161280} \right) L^8 
\notag \\ & -     
      a^5 \left( \frac{C_F^4 C_{A}}{576}  
 + \frac{ 19 C_F^3 C_{A}^2}{1024} + \frac{251 C_F^2 C_{A}^3}{967680}
 + \frac{151 C_F C_{A}^4}{5806080}\right) L^{10} \notag \,,\\
P_5 \; & = \; 
       \dfrac{1}{5!}\left(\dfrac{a C_F L^2}{2}\right)^5 \exp{\left[ -\frac{a C_F L^2}{2} \right]}  
          +  a^5 \left( \frac{C_F^4 C_{A}}{2304}  
 + \frac{ 13 C_F^3 C_{A}^2}{46080} + \frac{79 C_F^2 C_{A}^3}{967680}
 + \frac{31 C_F C_{A}^4}{3628800}\right) L^{10} \,,\\
P_{n\ge 6} & = \; 
       \dfrac{1}{n!}\left(\dfrac{a C_F L^2}{2}\right)^n \exp{\left[ -\frac{a C_F L^2}{2} \right]} \,.        
\end{alignat*}
}

\newpage


\end{document}